\documentclass[twoside,twocolumn]{article}
\usepackage{geometry}
\usepackage{setspace}
\usepackage{amsmath}
\usepackage{amssymb}
\usepackage{authblk}
\usepackage{float}
\usepackage{graphicx}
\usepackage{hyperref}
\usepackage[version=4]{mhchem}
\usepackage{bm}

\usepackage[backend=biber, style=numeric-comp, sorting=none]{biblatex}
\title{Large barrier behaviour of the rate constant from the diffusion equation}
\author[1]{Pierpaolo Pravatto}
\author[1]{Barbara Fresch}
\author[1]{Giorgio J.Moro}
\affil[1]{Dipartimento di Scienze Chimiche, Università di Padova, via Marzolo 1 - 35131 Padova (Italy)}

\date{}
\begin{document}
\twocolumn[
    \begin{@twocolumnfalse}
        \maketitle
        \begin{abstract}
        Many processes in chemistry, physics, and biology depend on thermally activated events in which the system changes its state by surmounting an activation barrier. Examples range from chemical reactions, protein folding, and nucleation events. 
           Parameterized forms of the mean-field potential are often employed in the stochastic modeling  of activated processes. In this contribution, we explore the alternative of employing parameterized forms of the equilibrium distribution by means of the symmetric linear combination of two gaussian functions. Such a procedure leads to flexible and convenient models for the landscape and the energy barrier whose features are controlled by the second moments of the gaussian functions. The rate constants are examined through the solution of the corresponding diffusion problem, that is the Fokker-Planck-Smoluchowski equation specified according to the parameterized equilibrium distribution. The numerical calculations clearly show that the asymptotic limit of large barriers does not agree with the results of the Kramers theory. The underlying reason is that the linear scaling of the potential, the procedure justifying the Kramers theory, cannot be applied when dealing with parameterized forms of the equilibrium distribution. A different kind of asymptotic analysis is then required and we introduce the appropriate theory when the  equilibrium distribution is represented as a symmetric linear combination of two gaussian functions, first in the one-dimensional case and afterward in the multi-dimensional diffusion model. 
            \\
        \end{abstract}
    \end{@twocolumnfalse}
]

\section{Introduction}
Arrhenius law~\cite{Arrhenius1889} is a fundamental tool of chemical kinetics~\cite{Laidler1987} and for the analysis of the rates of activated processes~\cite{Hanggi1990}. It allows their simple interpretation as dynamical processes controlled by the crossing of the energy barrier separating reactant and product states, like in the Transition State Theory~\cite{Eyring1935, Wigner1938}. Kramers in his article~\cite{Kramers1940} of 1940 proposed the stochastic description as the convenient scheme for treating the relaxation to equilibrium in activated processes through the frictional coupling with the environment. In this framework, he introduced the simplest and most insightful model for kinetics  according to the one-dimensional diffusion equation in the presence of a mean-field potential as described by the Fokker-Planck-Smoluchowski (FPS) equation~\cite{Kampen2007, Risken1989, Gardiner1985}. In particular, he showed that the Arrhenius law is recovered in the limit of large barriers, with the pre-exponential factor controlled by the diffusion coefficient and the second derivative (curvature) of the mean-field potential at the saddle point.

This Kramers Asymptotic Relation (KAR) has been highly appreciated in the literature and nowadays it appears as a standard tool in Physics, Chemistry, and Biophysics~\cite{Peters2017, Elber2020}. An important methodological issue then arises: does KAR uniquely identify the large barrier behavior of the one-dimensional diffusion model? In all generality, the answer is negative, and already Kramers~\cite{Kramers1940} provided a counterexample for the case of the mean-field potential with an edge shape barrier as obtained by matching two parabolic wells. Correspondingly the second derivative of the potential at the top of the barrier is not defined and the standard KAR is out of the question.

Still one might speculate that KAR is valid for the physically more realistic potential of having smooth behavior in all its derivatives. In this work, we intend to analyze the validity of KAR within this more restricted kind of model. We shall show that even in the presence of a smooth barrier, the asymptotic behavior of the kinetic rate is not necessarily described by KAR. An enlightening example is provided by the mean-field potential corresponding to the Two-Gaussian Distribution (TGD), which is the symmetric linear combination of two normal distributions having the same width for a given separation of their centers. The TGD was recently employed for the analysis of the tunneling splitting in light of the isomorphism between FPS operator and the Hamiltonian operator of Quantum Mechanics~\cite{Pravatto2022}. It might be considered also as a useful tool for modeling activated processes, as long as the corresponding mean field potential displays a two-well profile separated by a smooth barrier whose height depends on the unique control parameter given by the ratio between the width and the separation of the gaussians. Indeed, larger potential barriers are recovered by decreasing the superposition between the two gaussians, that is by decreasing their width or increasing their separation. We think that the mean-field potential of TGD is a convenient candidate for a parameterized form to be employed in the characterization of activated processes like other standard models, for instance, polynomial potentials~\cite{Risken1989, Larson1978, Shizgal1984, Moro1995}.

The spectral analysis of the FPS operator can be performed numerically to obtain the exact value of the rate constant for an increasing barrier height  $\Delta U$. This kind of computation shows that the asymptotic limit of TGD model for large barriers $\Delta U$ is not described by KAR, more precisely that it requires a different pre-exponential factor in the Arrhenius representation. This provides clear evidence that KAR has no general validity even in the case of smooth potentials.

The issue has its origin in the very definition of the asymptotic limit with respect to the barrier height $\Delta U$. The analysis of the kinetic rate by increasing the barrier requires a change of the mean-field potential and this can be done in different ways according to the parameterization of the stochastic model. Correspondingly different asymptotic relations might be recovered for the rate constant.

The simplest asymptotic procedure is certainly that of scaling the mean-field potential by multiplying it, as well as the potential barrier,  by a scaling factor so leaving unmodified the overall shape of the potential, as conveniently done for instance with polynomial potentials. This is considered in the literature as the standard procedure leading to KAR in the limit of large barriers ~\cite{Peters2017, Elber2020, Hanggi1990, Berglund_2013}. However it  cannot be applied to TGD model as long as the change of the control parameter, i.e., the ratio between the width and the separation of the gaussians, cannot be accounted by a scaling factor of the mean-field potential. Thus the TGD model calls for a different kind of asymptotic analysis that we present in this work and that leads to a pre-exponential factor of the Arrhenius form differing from that of KAR by  $\sqrt{\pi /2}$, that is an increase of about 25\% on the rate. The peculiarity of the asymptotic behavior of TGD model is revealed by the difference not only  on the pre-exponential factor but also  on the profile of the kinetic eigenmode  of the diffusion (FPS) equation, that is the eigenfunction of the evolution  operator that describes the relaxation according to the kinetic rate. It should be mentioned that the kinetic eigenmode is strictly related to the committor function that in recent times has been recognized as a fundamental tool for describing the configuration dependence of rare events on the basis of the probability that a trajectory starting at a given point reaches the product state without visiting the reactant state~\cite{Bolhuis2002, Weinan2010, Peters2016, Berezhkovskii2019, Roux2022}. While the error function profile is recovered from the potential scaling leading to KAR, the asymptotic analysis of TGD leads to a rather different kinetic eigenmode as specified by the integral of the inverse of the hyperbolic cosine.

One-dimensional diffusion models are appealing for the simplicity of their analysis, but a more profound and realistic interpretation of activated processes requires multi-dimensional diffusion representations in order to leave room for the coupling between the reactive and non-reactive coordinates. The extension of the Kramers asymptotic analysis to multi-dimensional diffusion has been performed by Langer~\cite{Langer1969} who derived a relation, in the following Kramers-Langer Asymptotic Relation (KLAR), that represents the multi-dimensional generalization of KAR. On the other hand, multi-dimensional TGD can be easily formulated on the basis of normal distributions and they generate a family of mean field potentials that are suitable to describe bistability and activated processes. We shall show that a similar scenario is recovered by moving from one-dimensional to multi-dimensional problems. In particular, like KAR for one-dimensional models, KLAR does not describe the large barrier limit of TGD models which require a different kind of asymptotic analysis.

The article is organized as follows. In the next section, the one-dimensional TGD model is introduced and the exact numerical results for the rate constant are compared with the Kramers asymptotic relation for increasing barrier so providing evidence of the failure of KAR for this particular diffusion model. In the third section, we explain why the Kramers method should not be applied to one-dimensional TGD potential and we introduce the appropriate asymptotic analysis which explains the numerical results for such a model. In the following section, we report the generalization of these results to multidimensional problems. Thus we introduce the diffusion model with the multidimensional Two Gaussian Distribution (mTGD) and we present its asymptotic analysis based on the scaling of the widths of the gaussians. In order to verify such an asymptotic analysis we report also the comparison with the numerical results for a two-dimensional realization of mTGD model. In the final section, we summarize the results of our analysis.

\section{Rate constant from one-dimensional diffusion}

The one-dimensional diffusion in the presence of a mean-field potential is described by the following Fokker-Planck-Smoluchowski (FPS) equation~\cite{Kampen2007, Risken1989, Gardiner1985} for the time-dependent probability density $\rho _t (x)$ on the coordinate $x$ 
\begin{equation}\label{FPS}
    \frac{\partial}{\partial t} \rho _t (x) = - \hat \Gamma \rho _t (x),
\end{equation}
where $\hat \Gamma$ is the time evolution operator specified as
\begin{equation}\label{Gamma}
    \hat \Gamma= - \frac{\partial}{\partial x} D \rho _{eq}(x) \frac{\partial}{\partial x}\rho _{eq}(x)^{-1}
\end{equation}
and $D$ is the diffusion coefficient. The equilibrium distribution $\rho _{eq}(x)$ determines the infinite time limit of a generic probability density $\rho _t (x)$ and it can be specified according to the mean field potential $U(x)$ scaled by the thermal energy factor $k_BT$
\begin{equation}\label{rhoeq}
    \rho _{eq}(x)=\frac{e^{-U(x)}}{Z}=\lim_{t\to \infty}\rho _t (x),
\end{equation}
where $Z$ is the suitable normalization. In this work we consider  bistable models with symmetric potentials $U(-x)=U(x)$ having a saddle point at $x=0$ and two equivalent minima at $x=\pm x_0$. The energy barrier height, once scaled by $k_BT$, is then given as $\Delta U:=U(0)-U(\pm x_0)$. 

FPS eq.~\ref{FPS} in the presence of large enough barriers displays two different kinds of motions: the activated transitions between the two potential wells and the intrawell dynamics~\cite{Hanggi1990}. A formal description of these motions is recovered according to the eigenmodes  $\phi _n (x)$  of FPS equation defined according to the eigenfunctions of $\hat \Gamma$: 
\begin{equation}\label{eiegenfunctions}
    \hat \Gamma \rho _{eq}(x) \phi _n(x) =\lambda _n \rho _{eq}(x) \phi _n(x)
\end{equation}
for $n=0,1,2,\cdots$, with $\lambda _n\le \lambda _{n+1}$. Notice that eigenmode $\phi _0 (x)=1$ with $\lambda _0=0$ is associated to the stationary solution of eq.~\ref{FPS}, that is  
 the equilibrium distribution. The ensemble of these eigenmodes allows the representation of the generic distribution $\rho _t(x)$ as their linear combination  weighted by $\rho _{eq}(x)$, with coefficients depending on the time as $e^{-\lambda _nt}$, eigenvalue $\lambda _n$ being the relaxation rate of eigenmode $\phi _n (x)$ (see Section A in Supporting Material for details).
 
For large enough barriers $\Delta U$, the kinetic behavior emerges because of the spectral gap  $\lambda _1 << \lambda _2,\lambda _3, \cdots $, which differentiates the relaxation rate $\lambda _1$ for the transitions between the two potential wells and the ensemble of eigenvalues $\lambda _n$ for $n\ge 2$ describing the relaxation rates of intrawell dynamics~\cite{Larson1978, Moro1995}. Correspondingly one can identify the time scales of the model: $\tau _{kin}:=1/\lambda _1$ for the time scale of transition kinetics overcoming the potential barrier and  $\tau _{lr}:=1/\lambda _2$ for the time scale of local relaxation ($lr$) within a potential well. For long enough times to ensure a local equilibration about each potential well, that is for $t>>\tau _{lr}$, only the first two eigenmodes $\phi_0=1$ and $\phi_1(x)$ are required to specify the time evolution of the probability density:
\begin{equation}\label{regime cinetico}
    \rho_t(x)\simeq \rho _{eq}(x) + \langle\phi_1  \rvert \rho_0\rangle e^{-\lambda _1t} \phi_1(x)\rho _{eq}(x),
\end{equation} 
with a single exponential decay along $\phi _1 (x)$ which identifies the kinetic eigenmode.

On the other hand, also the reversible unimolecolar mechanism, $\ce{A ->[k] B}$, $\ce{B ->[k] A}$,  of chemical kinetics~\cite{Laidler1987} leads to a single exponential decay $e^{-2kt}$ of the concentrations. Thus FPS equation can be considered as a stochastic model explaining the kinetic behavior in terms of the diffusion dynamics at times longer than those required by the local equilibration, under the constraint of the time scale separation $\tau _{kin}>>\tau _{lr}$ due to the spectral gap. By matching the two exponential decays, $e^{-2kt}=e^{-\lambda _1 t}$,  one gets a precise identification of the rate constant $k$ according to the first non vanishing eigenvalue of FPS equation
\begin{equation}\label{k}
    k=\lambda _1/2
\end{equation}
Such a relationship allows the exact calculation of the rate constant $k$ from the numerical diagonalization of FPS operator $\hat \Gamma$ once the potential $U(x)$  is chosen with a large enough barrier $\Delta U$ to ensure the spectral gap $\tau _{kin}>>\tau _{lr}$.

Surely Kramers~\cite{Kramers1940} was the first to consider the large barrier behavior of the one-dimensional diffusion model and he derived an asymptotic form, in the following KAR for Kramers Asymptotic Relation, that in our notation is specified as
\begin{equation}\label{KAR}
    k^\infty _K= \frac{D}{2\pi} \sqrt{U_0^{(2)}\left| U_S^{(2)} \right| }\  e^{-\Delta U}
\end{equation}
where $U_0^{(2)}$ and $U_s^{(2)}$ are  the second derivatives (curvatures) of the potential at the minima and the saddle point, respectively. The superscript in $ k^\infty _K$ refers to  the asymptotic validity of eq.~\ref{KAR} with respect to the infinite barrier $\Delta U$ limit. By taking into account that $\Delta U$ denotes the energy barrier scaled by the thermal factor $k_BT$, KAR has the same structure of the Arrhenius equation with a precise identification of the pre-exponential factor.  Therefore the Kramers result instantiates the heuristic value of the Arrhenius law by validating it in a model of dynamics even if of simple diffusional type.

The simplicity of one-dimensional diffusion models allows their spectral analysis numerically and, therefore, the exact calculation of the rate constant $k$ according to eq.~\ref{k}. Then the asymptotic values $k^\infty _K$ of eq.~\ref{KAR} can be compared to the exact rate constants $k$ in order 
to verify the validity of KAR by looking at the convergence of $k^\infty _K$ to $k$ for an increasing barrier $\Delta U$. This procedure will be applied to two different kinds of mean field potentials. As the reference case for which the convergence of KAR has been already verified in the past~\cite{Larson1978, Moro1995}, we consider the quartic polynomial potential that is conveniently specified  as
\begin{equation}\label{quartic}
    U(x)=\Delta U [(x/x_0)^2-1]^2.
\end{equation}
Furthermore, we examine the potential due to the Two-Gaussian Distribution (TGD) for the equilibrium probability density~\cite{Pravatto2022}
\begin{equation}\label{distribuzione TGD}
    \begin{split}
    \rho _{eq}(x): & =\frac{1}{2} \mathcal{N}(x|x_0,\sigma^2) + \frac{1}{2} \mathcal{N}(x|-x_0,\sigma^2) = \\ & =\frac{e^{-(x-x_0)^2/2\sigma^2} + e^{-(x+x_0)^2/2\sigma^2}} {\sqrt{8\pi \sigma ^2}},
    \end{split}
\end{equation}
that is the symmetric linear combination of two normal distributions centered at $\pm x_0$ and having the same width $\sigma$. If $x_0$ is used as units of length, such a model has a unique control parameter specified as $\sigma /x_0$. We shall consider the situation of well-separated gaussians so that the maxima of $\rho _{eq} (x)$ are nearly coincident with their centers $\pm x_0$, with deviations  of the order of the superposition parameter $S:=e^{-2x_0^2/\sigma ^2}$~\cite{Pravatto2022}. In particular, we consider the control parameter in the range $0<\sigma / x_0<1/2$ which ensures a negligible superposition parameter:  $S<10^{-3}$. The corresponding mean field potential is defined as~\cite{Pravatto2022}
\begin{equation}\label{TGD potential}
     \begin{split}
    U(x)&=-\ln \left[\sqrt{8\pi \sigma ^2} \rho _{eq} (x)\right]= \\ &=
     \frac{x_0^2+x^2}{2\sigma ^2} -\ln \left[2 \cosh (x x_0/\sigma ^2)\right], 
     \end{split}
\end{equation}
with the following barrier height by neglecting contributions of the order of $S$
\begin{equation}\label{barrier height}
    \Delta U = \frac{x_0^2}{2\sigma ^2}-\ln 2.
\end{equation}
 Clearly, the previous equation allows the identification of the control parameter $\sigma / x_0$ on the basis of a chosen value of the barrier $\Delta U$.

\begin{figure}[ht!]
    \centering
    \includegraphics[width=0.95\columnwidth]{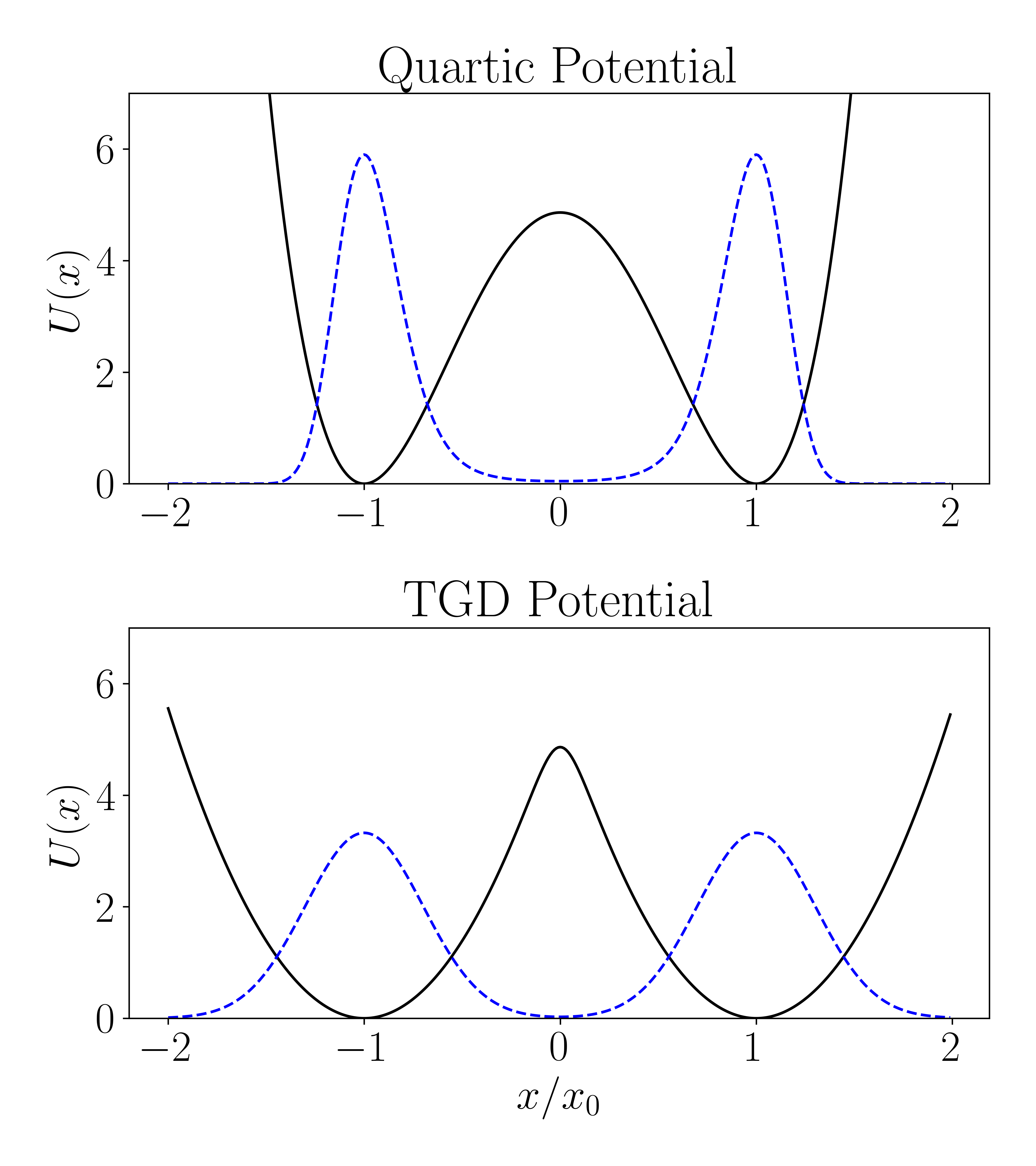}
    \caption{Mean field potential $U(x)$ (continuous line) and equilibrium distribution $\rho _{eq}(x)$ (dashed line) for $\Delta U\simeq4.86$. Upper panel: quartic potential of eq.~\ref{quartic}. Lower panel: TGD potential of eq.~\ref{TGD potential}}
    \label{fig:Figure1}
\end{figure}

In fig.~\ref{fig:Figure1} we have represented the mean-field potential together with the associated equilibrium distribution for both potentials with the same energy barrier of $\Delta U =4.86$, corresponding to a control parameter $\sigma/x_0=0.3$ for TGD model. It should be evident that TGD potential provides a reasonable profile  for bistable  symmetric problems and that it can be employed as an alternative to the quartic potential.

\begin{figure}[ht!]
    \centering
    \includegraphics[width=0.95\columnwidth]{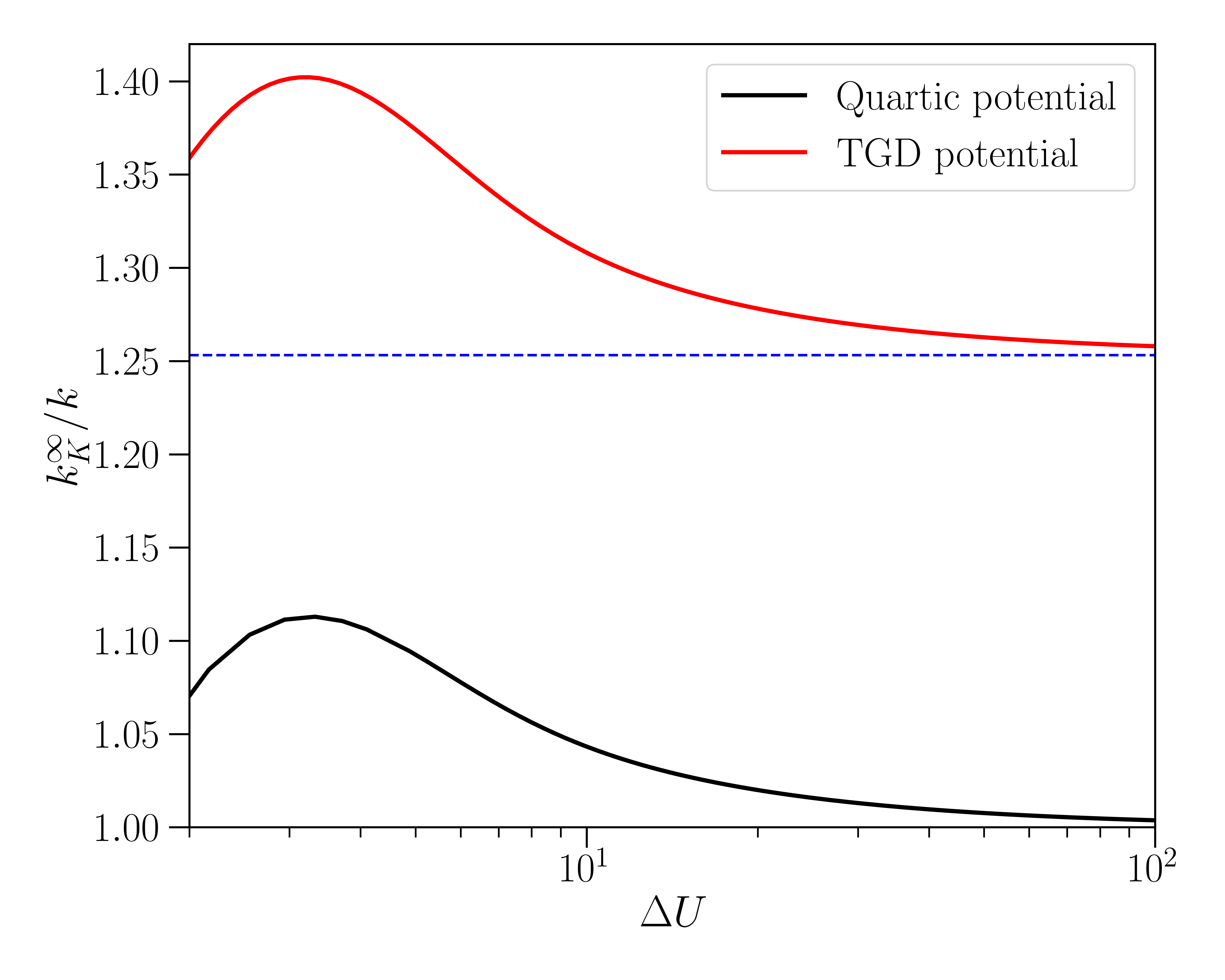}
    \caption{Ratio $k^\infty _K /k$ between  asymptotic Kramers relation (KAR) and the exact rate constant $k$  for increasing barrier height $\Delta U$. Black line: quartic potential; red line: TGD potential.}
    \label{fig:Figure2}
\end{figure}

We have evaluated the exact rate constant $k$ according to the first non-vanishing eigenvalue $\lambda _1$ computed  numerically  with both  potentials. The comparison with KAR is made in figure~\ref{fig:Figure2} in the form of the ratio $k^\infty _K/k$ as a function of the barrier height $\Delta U$. In the case of the quartic polynomial potential (black line in fig.~\ref{fig:Figure2}), the convergence to the asymptotic result of Kramers is detected  by the smooth approach to unity of the ratio $k^\infty _K /k$ with  increasing barrier $\Delta U$. A completely different behavior is recovered from the TGD potential (red line in fig.~\ref{fig:Figure2}) with, for large values of $\Delta U$, $k^\infty _K /k$ approaching a constant value different from unity like the one represented by the dashed line in fig.~\ref{fig:Figure2}. This represents  clear evidence that KAR eq.~\ref{KAR} does not describe correctly the asymptotic behavior in the case of TGD potential. In the next section, we analyze the reason for such a failure of the asymptotic Kramers relation.

\section{Asymptotic behavior of one-dimensional diffusion}\label{sec: saddle-point-expansion}

Let us first recall the procedure employed by Kramers in his 1940's article~\cite{Kramers1940}. He examined the stationary solution of the one-dimensional diffusion (FPS) equation by imposing source and sink boundary conditions at the potential minima. Then the rate constant was evaluated according to the flux-over-population ratio. This procedure is commonly adopted in the literature concerning the asymptotic limit of the rate constant for large barriers~\cite{Hanggi1990, Reimann1999, Bartsch2008}. Here we employ the alternative procedure of an asymptotic analysis performed directly on the first non-vanishing eigenvalue $\lambda _1$ of FPS operator, which according to eq.~\ref{k} ensures the exact identification of the rate constant. In this way, results equivalent to those of Kramers are recovered without modifications of the diffusion model with ad hoc boundary conditions.

The central property to examine is the kinetic eigenmode $g(x)\equiv \phi_1(x)$, which is the eigenfunction of operator eq.~\ref{Gamma} that describes the barrier crossing process, and which we specifically identify with the symbol $g(x)$~\cite{Pravatto2022}. It has a step-like profile concentrated in a narrow domain about the saddle point, while far from the saddle point is nearly constant with about $\pm 1$ values~\cite{Moro1995}. We recall that by a suitable scaling, one recovers from $g(x)$ the committor function~\cite{Peters2016, Roux2022}.  Because of the spectral gap $\lambda _1/\lambda _2 \to 0$ for $\Delta U\to \infty$, in the scale of the typical eigenvalue of FPS equation, say $\lambda _2$, the contribution of $\lambda _1$ vanishes and the equation providing the kinetic component $g(x)$ is approximated as $\hat \Gamma g(x)\rho _{eq}(x)=0$ which is equivalent to
\begin{equation}\label{g'(x)}
    \frac{\partial}{\partial x}\rho _{eq}(x) g'(x)=0,
\end{equation}
where $g'(x):=dg(x)/dx$ is non-vanishing only in a small range about the saddle point. This motivates the replacement of the the potential in $\rho _{eq}(x)\propto \exp \{ -U(x)\} $ with its second-order Taylor expansion
\begin{equation}\label{U parabolic}
    U(x) \simeq U(0)-\frac{1}{2} \left|U_s^{(2)} \right| x^2=U(0)-{x_K^\infty}^2,
\end{equation}
where
\begin{equation}\label{xK}
    x_K^\infty := x\sqrt {\left|U_s^{(2)}\right|/2} 
\end{equation}
is the scaled coordinate according to the asymptotic method of Kramers. The parabolic approximation  eq.~\ref{U parabolic} of the mean-field potential is the key ingredient of the Kramers asymptotic analysis. Correspondingly the solution of eq.~\ref{g'(x)} with the proper symmetry, $g(-x)=-g(x)$, and the boundary conditions $\lim _{x\to \pm \infty}g(x)=\pm 1 $ is readily found according to the error function
\begin{equation}\label{erf}
   g_K^\infty=\text{erf} ( x_K^\infty). 
\end{equation}
Notice the universal character of this Kramers asymptotic form of the kinetic eigenmode, as long as the parametric dependence on the diffusion model is taken into account only through the definition of the scaled coordinate $x_K^\infty$. Finally, by evaluating $\lambda _1$ as the expectation value of FPS operator according to the Kramers asymptotic form eq.~\ref{erf} of the kinetic eigenmode, KAR eq.~\ref{KAR} is recovered (see Section B of Supporting Material for details).

In order to understand why KAR eq.~\ref{KAR} works well in some cases but fails in others like TGD potential, a deeper analysis of the asymptotic procedure is required. Let us first stress that in all generality the first non-vanishing eigenvalue $\lambda _1$ and, therefore, the rate constant $k$ according to eq.~\ref{k}, is proportional to diffusion coefficient $D$ but depends on the detailed shape of the mean-field potential $U(x)$. This is specified in a formal sense as
\begin{equation}\label{k functional}
   k=D\ F[U(x)], 
\end{equation}
where $F[U(x)]$ denotes the functional dependence on $U(x)$. From this point of view, the numerical calculation of $\lambda _1=2k$ is a procedure for determining the value of the functional $F[U(x)]$ for a given potential $U(x)$. On the other hand, the asymptotic analysis with respect to the barrier height $\Delta U$ requires that the rate constant should be represented as an ordinary function of $\Delta U$
\begin{equation}\label{k function}
   k=D\ f(\Delta U),
\end{equation}
so to be in the position to extract from $f(\Delta U)$ the leading contribution for $\Delta U \to \infty$. In conclusion, a preliminary step of the asymptotic analysis is the conversion of the functional $F[U(x)]$ into a function $f(\Delta U)$ of the barrier height and this can be done in different ways according to the kind of mean-field potential we are dealing with.  Correspondingly different asymptotic relations could be found for the rate constant and,  in the following, we discuss two cases that exemplify the difference in the asymptotic predictions.

Let us consider first the scaling of the potential by the barrier, which can be considered the standard method leading to KAR~\cite{Berglund_2013}. It is based on the idea of multiplying the potential by a scalar factor that increases the barrier without modifying the overall potential profile.  It can be applied to models of potential which can be represented as 
\begin{equation}\label{shape}
  U(x)=\Delta U \ s(x), 
\end{equation}
where $s(x)$ is the potential shape function corresponding to the mean field potential with a unitary barrier. The quartic potential eq.~\ref{quartic}   belongs to this category with the shape function given as  $s(x)=[(x/x_0)^2-1]^2$. Then, by definition, the functional $F[U(x)]$ becomes a function $f(\Delta U)$ of the potential barrier only,  if we consider the class of potentials eq.~\ref{shape} for a fixed potential shape $s(x)$.  In this framework, one can justify the truncation at the second order eq.~\ref{U parabolic} of the Kramers derivation because of the asymptotic limit $\Delta U \to \infty$. Indeed, according to the Taylor expansion of the potential eq.~\ref{shape} for $x\to 0$
\begin{equation}\label{Taylor shape}
    \begin{split}
  U(x)&=U(0) +\Delta U s_s^{(2)}x^2/2+\Delta U O(x^4)= \\ &= U(0) - {x^\infty_K}^2+ \frac {1} {\Delta U}O({x^\infty_K}^4), 
      \end{split}
\end{equation}
where $ x^\infty_K= x \sqrt{\Delta U d^2s(x)/dx^2|_{x=0} /2}$, the limit $\Delta U \to \infty$ ensures the validity of the parabolic form eq.~\ref{U parabolic}, because the fourth and higher order terms become negligible, as well as the validity of the Kramers Asymptotic Relation eq.~\ref{KAR}. The convergence to the unity of $k^\infty _K/k$ for increasing $\Delta U$ as displayed in Fig.~\ref{fig:Figure2} for the quartic potential, is strictly a consequence of this asymptotic behavior of  potentials parameterized as in eq.~\ref{shape}.

TGD potential eq.~\ref{TGD potential} cannot be assimilated to the form eq.~\ref{shape} since it has $\sigma /x_0$ as the only control parameter whose changes are not reproduced by a simple scaling of the potential. In other words, the scaling of eq.~\ref{TGD potential} by a multiplicative factor does not define a mean-field potential that can be generated by the Two-Gaussian-Distribution  eq.~\ref{distribuzione TGD}. Therefore, the previous asymptotic analysis cannot be applied to TGD potential. On the other hand, the preliminary step of the reduction of the functional dependence of the rate constant from eq.~\ref{k functional} to eq.~\ref{k function} is not required with TGD potential because it is parameterized according to the control parameter $\sigma /x_0$ that  can be specified by the barrier height $\Delta U$ because of eq.~\ref{barrier height}. This implies the equivalence of the asymptotic limit $\Delta U \to \infty$ with a vanishing control parameter $\sigma /x_0\to 0$. In conclusion, the functional dependence of the rate constant of TGD model is specified by eq.~\ref{k function} by construction. Furthermore one can easily show that the parabolic form eq.~\ref{U parabolic} does not hold asymptotically in the case of TGD potential, since for $x\to 0$
\begin{equation}\label{TGD parabolic}
    \begin{split}
    U(x) & = U(0)-\frac{x^2}{2\sigma ^2} \left(  \frac{x_0^2}{\sigma ^2} - 1 \right) + O (x^4 x_0^4/\sigma ^8 )= \\
    & =U(0)-{x_K^\infty}^2 + O({x_K^\infty}^4),
      \end{split}
\end{equation}
where $x_K^\infty =x\sqrt{(x_0^2-\sigma^2)/2\sigma^4}$ is the Kramers scaled coordinate evaluated according to  eq.~\ref{xK}. It is clear that in the limit $\sigma/x_0 \to 0$ , the fourth and higher order terms are not negligible unlike in eq.~\ref{Taylor shape}. Therefore, the asymptotic parabolic form eq.~\ref{U parabolic} of the potential, which is the basic ingredient of the Kramers procedure, does not find justification in the case of TGD potential.

A different procedure has to be developed in order to find the asymptotic behavior of the rate constant evaluated with TGD potential. One can easily recognize that in the TGD potential eq.~\ref{TGD potential} the hyperbolic cosine term prevails in the limit $\sigma/x_0 \to 0$ and this suggests taking its argument as the scaled variable of the asymptotic procedure
\begin{equation}\label{xTGD}
  x^\infty_{TGD}:=xx_0/\sigma ^2, 
\end{equation}
where we have inserted the subscript TGD to emphasize the reference to such a model potential. By using $x^\infty_{TGD}$ as the independent variable, TGD potential eq.~\ref{TGD potential} is rewritten as
\begin{equation}\label{TGD potential 2}
  U(x)-U(0)= \frac{\sigma^2}{2x_0^2} 
  {x^\infty_{TGD}}^2 -\ln \left[\cosh\left(x^\infty_{TGD}\right) \right],
\end{equation}
so that, in the asymptotic limit $\sigma /x_0 \to 0$, only the second term at r.h.s. survives. Thus, instead of the gaussian distribution  $\rho _{eq}(x)\propto \exp \{ -U(x) \} )\propto \exp (-{x^\infty _K}^2 )$ derived with the Kramers procedure, a different asymptotic behavior specified by the hyperbolic cosine is recovered for TGD potential 
\begin{equation}\label{rho eq TGD}
  \rho _{eq} (x) \propto \cosh \left( x^\infty_{TGD} \right).
\end{equation}
Correspondingly, by integrating eq.~\ref{g'(x)}, the following form is recovered for the kinetic eigenfunction of FPS,
\begin{equation}\label{g TGD}
  g^\infty_{TGD}=\frac{2}{\pi} \int _0 ^{ x^\infty_{TGD}} \frac{dy}{\cosh(y)},
\end{equation}
which replaces the result eq.~\ref{erf} of the Kramers procedure in the case of TGD potential. Notice that the proportionality coefficient in the previous equation has been chosen in order to ensure the limiting behavior $\lim _{x\to \pm \infty}g^\infty_{TGD}(x)=\pm 1$. Eq.~\ref{g TGD}, like eq.~\ref{erf}, has the structure of a universal form of the kinetic eigenfunction  since it does not bear any reference to the parametric dependence of the model, which is taken into account only through the definition of the scaled variable, $x^\infty_{TGD}$ and $x^\infty_K$ in the two cases.

Eq.~\ref{g TGD} for the kinetic eigenmode  of FPS equation is the main result of our asymptotic analysis of TGD potential. From the corresponding expectation value of FPS operator (see Section B of Supporting Material for details) the following asymptotic rate constant is derived
\begin{equation}\label{k TGD}
  k^\infty _{TGD}=D\frac{x_0}{\sigma ^3 \sqrt{2\pi ^3}} \ e^{-\Delta U},
\end{equation}
again of Arrhenius type with a pre-exponential factor determined by the control parameter. One might wonder what would be the result of the KAR  eq.~\ref{KAR} if applied to TGD potential. By inserting into eq.~\ref{KAR} the second derivatives of potential eq.~\ref{TGD potential}, in the asymptotic limit $\sigma /x_0 \to 0$ one obtains a pre-exponential factor different from that of eq.~\ref{k TGD} by a factor $\sqrt{\pi /2}$ 
\begin{equation}\label{ratio k}
  \lim _{\sigma/x_0\to 0} \frac{k^\infty_K}{k^\infty _{TGD}}= \sqrt{\pi /2}.
\end{equation}
This explains the results for the ratio ${k^\infty_K}/k$ for TGD potential as displayed by the red line in Fig.~\ref{fig:Figure2}. Such a ratio reaches the asymptotic value of $\sqrt{\pi /2}$ (the dashed line of Fig.~\ref{fig:Figure2}) just because the Kramers relation over-estimates by the same factor the asymptotic correct result eq.~\ref{k TGD} for TGD potential. On the other hand, the evidence from Fig.~\ref{fig:Figure2} that $\lim _{\sigma /x_0\to 0} k^\infty _K/k=\sqrt{\pi /2}$ for TGD potential  can be considered that the verification that the previous asymptotic analysis is correct.

To summarize, we have shown that when dealing with TGD model, an asymptotic analysis different from that proposed by Kramers~\cite{Kramers1940} is necessary and this leads to an asymptotic rate constant still of Arrhenius type but with a different pre-exponential factor. The difference, however, is not confined to the pre-exponential factor of Arrhenius form since a completely different asymptotic profile is derived for the kinetic eigenmode: the integral of the inverse hyperbolic cosine of  eq.~\ref{g TGD} versus the error function profile of eq.~\ref{erf}. As mentioned before, this implies a different behavior of the committor function~\cite{Bolhuis2002, Weinan2010, Peters2016, Berezhkovskii2019, Roux2022} in the two cases. To illustrate it, Figure 3 provides a comparison between these asymptotic forms $g^\infty _K$ of  eq.~\ref{erf} and $g^\infty _{TGD}$ of eq.~\ref{g TGD} as a function of $x^\infty_K$ and $x^\infty _{TGD}$, respectively. Their difference is evident in the approach to the asymptotes $\pm 1$.

\begin{figure}[ht!]
    \centering
    \includegraphics[width=0.95\columnwidth]{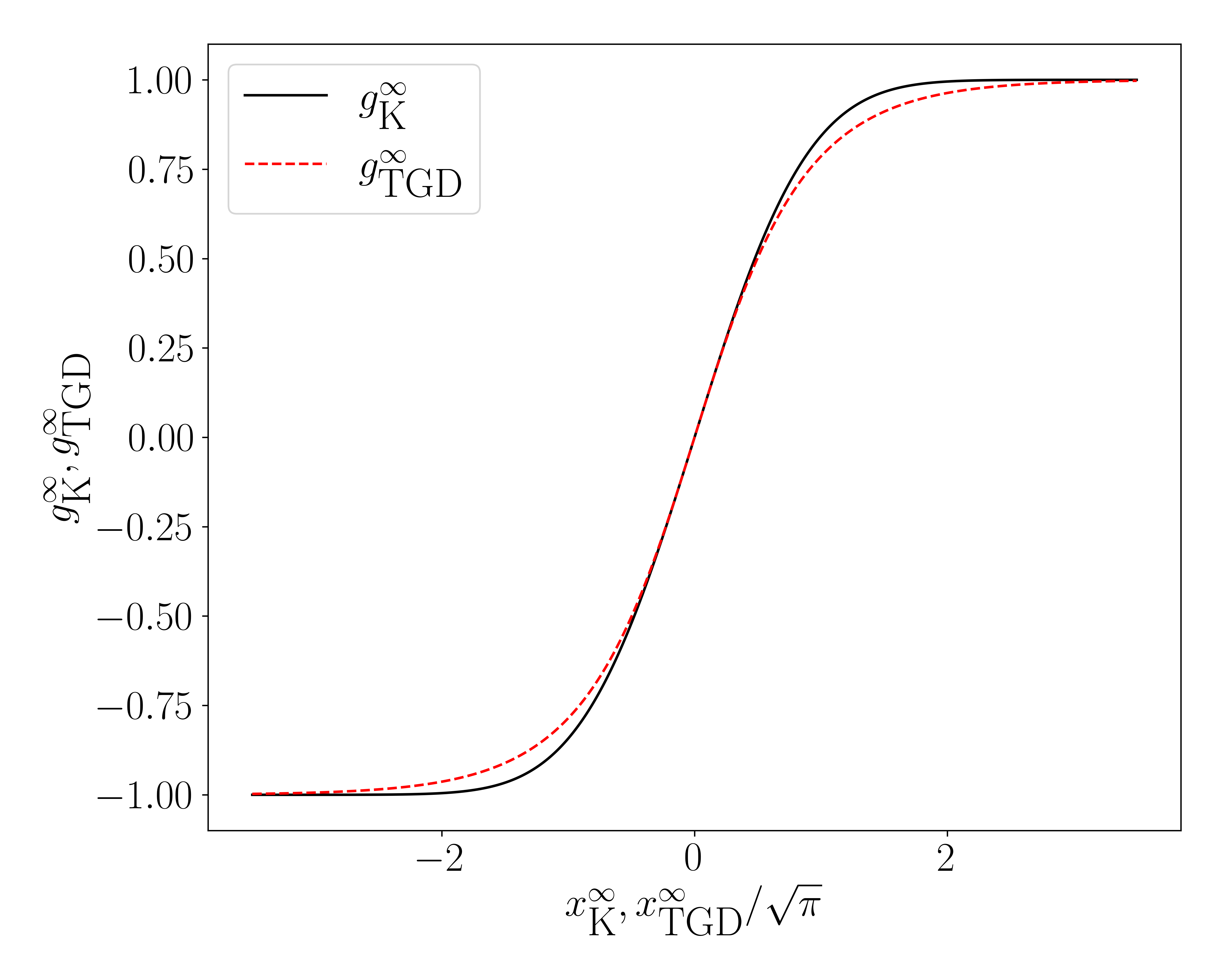}
    \caption{Asymptotic forms of the kinetic eigenfunction $g^\infty _K$ and  $g^\infty _{TGD}$  as function of $x^\infty_K$ and of $x^\infty _{TGD}$, respectively, with the latter scaled as  $x^\infty _{TGD}/\sqrt{\pi}$ in order to to ensure in the graph the same first derivative at the origin.}
    \label{fig:Figure3}
\end{figure}

\section{Asymptotic behavior of multidimensional diffusion}\label{sec: TGD-models-nD}

In this section, we intend to generalize the previous analysis to multidimensional diffusion models. Even if these systems are in general much more complex, still asymptotic behavior can be reduced to the same one-dimensional forms for  the kinetic eigenmode. Furthermore, it will be shown that the multidimensional Two Gaussian Distribution (mTGD) model is characterized by an asymptotic rate constant different from the Kramers-Langer Asymptotic Relation (KLAR), that is the multidimensional generalization by Langer~\cite{Langer1969} of the Kramers result. The presentation  is organized according to the following lines. First, we introduce the Fokker-Planck-Smoluchowski (FPS) description of diffusion in symmetric bistable multidimensional problems and the multidimensional Two Gaussian Distribution (mTGD). Second, we recall the Langer analysis~\cite{Langer1969} which, from the normal mode analysis at the saddle point, allows the one-dimensional reduction of the kinetic eigenmode to the form
eq.~\ref{erf} with a suitable definition of the asymptotic reaction coordinate. Afterward, we tackle the asymptotic analysis for mTGD potentials  leading to the kinetic eigenmode  of the same form of  eq.~\ref{g TGD}. On this basis, the comparison is made  between the asymptotic analysis of mTGD potential and the predictions of KLAR. Finally, the results of our analysis is validated by examing the numerically exact rate constant for a two-dimensional realization of the mTGD model.

\subsection{Multidimensional Two Gaussian (mTGD) model}
The $N$-dimensional diffusion problem for coordinates $\mathbf{x}=\{x_1, x_2, ..., x_N\}$ is described by the time-dependent probability density $\rho _t(\mathbf{x})$ normalized by integration on each coordinate in the full real axis and evolving in time as in eq.~\ref{FPS}  with the following Fokker-Planck-Smoluchowski operator~\cite{Kampen2007, Risken1989, Gardiner1985}
\begin{equation}\label{mGamma}
    \hat \Gamma= - \frac{\partial}{\partial \mathbf{x}}^T \mathbf{D} \rho _{eq}(\mathbf{x}) \frac{\partial}{\partial \mathbf{x}}\rho _{eq}(\mathbf{x})^{-1},
\end{equation}
where $\mathbf{D}$ is the $N\times N$ diffusion matrix supposed to be independent of the coordinates. Let the operator $\hat R$ encode the binary symmetry, $\hat R^2=\hat 1$, of bistable symmetric systems with an invariant equilibrium distribution and a commuting evolution operator
\begin{equation}\label{symmetry}
    \hat R \rho _{eq}(\mathbf{x})=\rho _{eq}(\mathbf{x}), \ \ \ \ \left[ \hat \Gamma, \hat R \right ]=0,
\end{equation}
implying the degeneracy of the properties and of the dynamics in the presence of an infinite barrier separating the two states. It is convenient to employ coordinates that are irreducible representations of the symmetry group, $(\hat R, \hat I)$, of the problem. Thus the coordinates are partitioned as $\mathbf{x}=(\mathbf{x}^+, \mathbf{x}^-)$
where $\mathbf{x}^+=(x_1^+,x_2^+, \cdots, x_{N^+}^+)$ and $\mathbf{x}^-=(x_1^-,x_2^-, \cdots, x_{N^-}^-)$ are even and odd, respectively, with respect to symmetry operator $\hat R$:
\begin{equation}\label{Rx}
    \hat R x^\pm _n=\pm  x^\pm _n.
\end{equation}
$N^+$ and $N^-$  denote the number of even coordinates and odd coordinates, respectively, with $N=N^++N^-$ being the overall number of coordinates.
Then the action of the symmetry operator $\hat R$
 on a function $f(\mathbf{x})$ of the coordinates is algebraically reduced to the multiplication of the array $\mathbf{x}$ by the square matrix $\mathbf{R}$,
 \begin{equation}\label{Rf(x)}
    \hat R f(\mathbf{x})=f(\mathbf{R} \mathbf{x}),
\end{equation}
with the matrix $\mathbf{R}$ block partitioned with respect to even and odd coordinates as 
\begin{equation}\label{Reflection_matrix}
    \mathbf{R}=
        \begin{pmatrix}
        \mathbf{R}^{++} & \mathbf{R}^{+-}\\
        \mathbf{R}^{-+} &\mathbf{R}^{--} 
    \end{pmatrix} =
    \begin{pmatrix}
        \mathbf{I}^+ & 0\\
        0 & -\mathbf{I}^-
    \end{pmatrix},
\end{equation}
where $\mathbf{I}^+$ and $\mathbf{I}^-$ represent the identity matrix for the set of even coordinates and odd coordinates, respectively. Then one easily derives that the commutation condition in eq.~\ref{symmetry} implies the constraint $\mathbf{R}\mathbf{D}\mathbf{R}=\mathbf{D}$, that is the vanishing of the off-diagonal blocks of the diffusion matrix partitioned like in eq.~\ref{Reflection_matrix} with respect to even and odd coordinates:
\begin{equation}\label{D}
    \mathbf{D}=
        \begin{pmatrix}
        \mathbf{D}^{++} & 0\\
       0 &\mathbf{D}^{--} 
    \end{pmatrix}. 
\end{equation}
Like for the one-dimensional case, a fundamental tool for the analysis of the rate constant is the kinetic eigenmode $g(\mathbf{x})$ which, in the limit of large barriers producing a large spectral gap, can be evaluated as the solution of the equation
\begin{equation}\label{g multidimensional}
    \frac{\partial}{\partial \mathbf{x}}^T \mathbf{D} \rho _{eq}(\mathbf{x}) \frac{\partial}{\partial \mathbf{x}} g(\mathbf{x})=0,
\end{equation}
that is the multi-dimensional generalization of eq.~\ref{g'(x)}. Once the asymptotic form $g^\infty(\mathbf{x})$ of the kinetic eigenfunction is determined, the rate constant can be evaluated according to the corresponding expectation value of FPS evolution operator eq.~\ref{mGamma}.

In the following, as an example of a system that does not follow KLAR, we shall specifically consider the multidimensional generalization (mTGD) of the Two Gaussian Distribution eq.~\ref{distribuzione TGD} specified as the symmetric linear combination of two Normal distributions,
\begin{equation}\label{mTGD}
 \begin{split}
    \rho &_{eq} (\mathbf{x}):  =\frac{1+\hat R}{2}  \mathcal{N}(\mathbf{x}|\mathbf{x}_0,\bm{\Sigma}) =
    \frac{1}{2 \sqrt{ (2\pi)^N \det({\bm{\Sigma}} ) } }
    \times \\
    & \times    \{ \exp[-(\mathbf{x}-\mathbf{x}_0)^T \bm{\Sigma}^{-1}(\mathbf{x}-\mathbf{x}_0)/2]
    + \\
    & \ \ \ \ + \exp [-(\mathbf{R}\mathbf{x}-\mathbf{x}_0)^T \bm{\Sigma}^{-1}(\mathbf{R}\mathbf{x}-\mathbf{x}_0)/2] \},  
   \end{split}
\end{equation}
characterized according to the positive definite matrix $\bm{\Sigma}$ of the second moments (covariance matrix) and their centers (first moments) at $\mathbf{x}_0=(\mathbf{x}_0^+, \mathbf{x}_0^-)$ and at $\mathbf{R}\mathbf{x}_0=(\mathbf{x}_0^+, -\mathbf{x}_0^-)$.  In the following the eigenvalues of the covariance matrix, $\mathbf{\Sigma}$, will be denoted by $\sigma _k^2>0$ for $k=1,2,\cdots,N$.
Since the origin of even coordinates $\mathbf{x}^+$ is arbitrary, we choose it in correspondence of the centers of the two gaussians: $\mathbf{x}^+_0=0$ so that $\mathbf{x}_0=(0, \mathbf{x}_0^-)$ and $\mathbf{R}\mathbf{x}_0= (0, -\mathbf{x}_0^-)= -\mathbf{x}_0$. Notice that the first moment cannot have a vanishing odd component  $ \mathbf{x}_0^-$ since otherwise the two gaussians would be centered  at the same location.
The mean field potential of mTGD is defined in analogy to eq.~\ref{TGD potential}
\begin{equation}\label{mTGD potential}
    U(\mathbf{x}) := -\ln \left[ 2 
    \sqrt{ (2\pi)^N \det({\bm{\Sigma}} ) }
    \ \rho _{eq} (\mathbf{x})\right].
\end{equation}
We assume a weak superposition between the two gaussians as quantified by negligible values of the superposition parameter $S:=\exp [ -2\mathbf{x}_0^T\bm{\Sigma}^{-1}\mathbf{x}_0]$, so that the minima of the potential are found at the centers of the gaussians, that is  at $\mathbf{x}_0$ and  $-\mathbf{x}_0$. The saddle point $\mathbf{x}_s=(\mathbf{x}_s^+,0)$ has vanishing odd coordinates by symmetry, while its even coordinates are  evaluated from the condition of vanishing gradient of the potential
\begin{equation}\label{saddle}
    \mathbf{x}_s^+=\left [ (\mathbf{\Sigma}^{-1})^{++}      \right]^{-1}(\mathbf{\Sigma}^{-1})^{+-}\mathbf{x}_0^-.
\end{equation}
By neglecting terms of the order of the superposition parameter $S$, the barrier height is given as: 
\begin{equation}\label{barrier mTGD}
    \Delta U = U(\mathbf{x}_s)-U(\mathbf{x}_0)=
    \frac{1}{2}     (\mathbf{x}_0^-)^T(\mathbf{\Sigma}^{--})^{-1}\mathbf{x}^-_0- 
    \ln(2).
\end{equation}
Furthermore, by employing the displacements $\delta\mathbf{x} := \mathbf{x} - \mathbf{x}_s$ from the saddle point, the potential eq.~\ref{mTGD potential} can be decomposed as 
\begin{equation}\label{mTGD potential 2}
    U(\mathbf{x}) = U_+(\delta\mathbf{x})- \ln \{ \cosh [U_-(\delta\mathbf{x})]  \}, 
\end{equation}
with the two components  $U_\pm$ with opposite symmetry, $\hat R U_\pm (\delta\mathbf{x})=\pm U_\pm (\delta\mathbf{x})$, given as
\begin{equation}\label{mTGD potential 3}
 \begin{split}
       & U_+(\delta \mathbf{x})=U(\mathbf{x}_s)+ \frac{1}{2} (\delta \mathbf{x}^+)^T (\mathbf{\Sigma}^{-1})^{++} \delta\mathbf{x}^+ +\\
        & \ \ \ \ \ \ \ \ \ \ \ \ +\frac{1}{2} (\delta \mathbf{x}^-)^T (\mathbf{\Sigma}^{-1})^{--} \delta\mathbf{x}^-\\
   & U_-(\delta\mathbf{x})= (\delta\mathbf{x}^-)^T(\mathbf{\Sigma}^{--})^{-1}  \mathbf{x}_0^- - \\
   & \ \ \ \ \ \ \ \ \ \ \ \ -(\delta \mathbf{x}^-)^T (\mathbf{\Sigma}^{-1})^{-+}\delta \mathbf{x}^+
\end{split}
\end{equation}
(for details see Section C in Supporting Material).

A kinetic interpretation can be attributed to the coordinates 
$\mathbf{x}=(\mathbf{x}^+, \mathbf{x}^-)$ on the basis of the critical points of the mean-field potential $U(\mathbf{x})$. If we associate the two minima $\mathbf{x}_0$ and $-\mathbf{x}_0$ to the two species of the kinetics, then the odd coordinates $\mathbf{x}^-$ assumes the character of collective coordinates of reaction as long as their values change in the transition. On the contrary, the even coordinates $\mathbf{x}^+$ can be interpreted as variables of non-reaction since they are not modified by a jump from one minimum to the other. However, the off-diagonal blocks of the variance matrix $\mathbf{\Sigma}$ introduce a coupling between these two kinds of coordinates which is responsible for the displacement $\mathbf{x}_s^+$ of the saddle point from the origin. In the absence of such a coupling for $\mathbf{\Sigma}^{+-}=\mathbf{\Sigma}^{-+}=0$, the saddle point is at the origin, that is midway between the two minima and the transition would not be affected by the displacement of even coordinates. On the contrary, for $\mathbf{\Sigma}^{+-} \neq 0$ the activated process is controlled also by the displacement of the coordinates of non-reaction.

\begin{figure}[ht!]
    \centering
    \includegraphics[width=0.95\columnwidth]{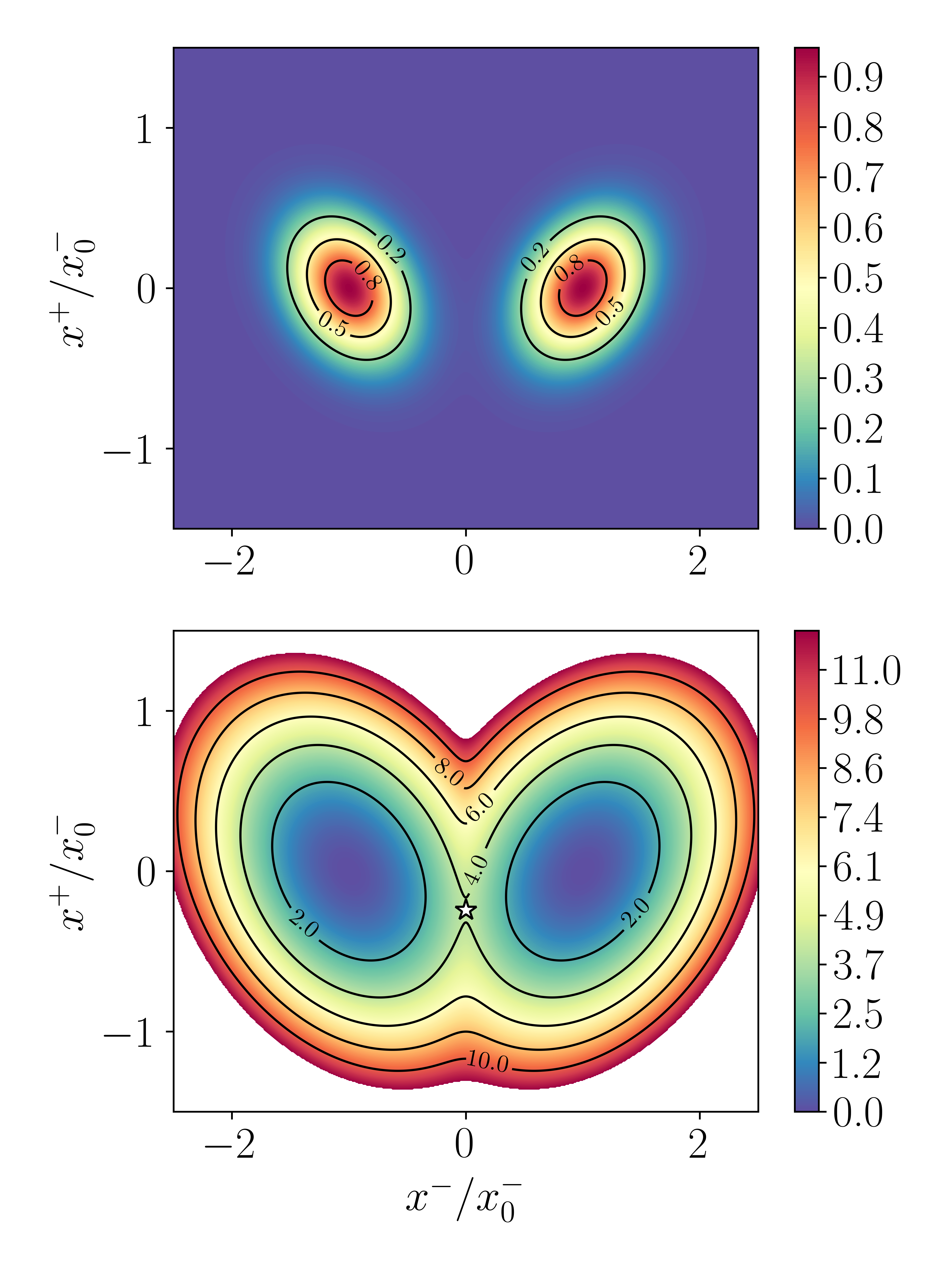}
    \caption{ 
    Color code representation of the equilibrium distribution $\rho _{eq}(\mathbf{x})$ (upper panel) and of the corresponding potential $U(\mathbf{x})$  (lower panel), for the two-dimensional mTGD model  with $\mathbf{x}=({x}^+, {x}^-)$ coordinates and parameters: $\theta=\pi/6,\  \sigma _1/x_0^-=0.25,\  \sigma _2/x_0^-=0.35 $. The star symbol in the lower panel denotes the position of the saddle point $\mathbf{x}_s$.}
        \label{fig: TGD_shape}
\end{figure}

As a simple example of mTGD we shall consider here and in the following the two-dimensional model with one coordinate of even and odd type, $\mathbf{x}=({x}^+, {x}^-)$. To parameterize the $2\times 2$ variance matrix $\mathbf{\Sigma}$ we employ its principal values, $\sigma _1^2$ and $\sigma _2^2$, and the angle $\theta$ between the principal direction for $\sigma _1^2$ and the axis of coordinate $x^+$. For $\theta =0$, $x^+$ and $x^-$ coordinates are independently distributed and also dynamically uncoupled since according eq.~\ref{D} the diffusion matrix is diagonal. In this case, the activated process is described by the diffusion of odd coordinate $x^-$ alone like in the analysis of the previous section. A non-vanishing angle, $\theta \neq 0$, for the variance matrix introduces a coupling between the two coordinates that requires an intrinsically two-dimensional description of the diffusion.  In Figure~\ref{fig: TGD_shape} we have represented  by means of a color code both the equilibrium distribution $\rho _{eq}(\mathbf{x})$ (upper panel) and the corresponding potential  $U(\mathbf{x})$ (lower panel) for the particular case described by parameters: $\theta=\pi/6,\  \sigma _1/x_0^-=0.25,\  \sigma _2/x_0^-=0.35 $. This is a situation of weakly superimposed gaussians, as clearly seen in the upper panel, which leads to a significant potential barrier of $\Delta U = 3.958$. In the lower panel for the potential, the displacement of the saddle point (star symbol) from the origin is also evident, as a consequence of the coupling  of the two coordinates induced by the variance matrix $\mathbf{\Sigma}$ for $\theta\neq 0$.

\subsection{Asymptotic analysis}
Before analyzing the asymptotic behavior of mTGD model, for the sake of comparison,  we recall the essential elements of the Kramers-Langer theory of the kinetic rate in multidimensional diffusion systems~\cite{Hanggi1990, Berglund_2013, Langer1969}. It is based on the second-order expansion of the potential at the saddle point
\begin{equation}\label{U(2)s}
    U(\mathbf{x})-U(\mathbf{x}_s) =  \delta \mathbf{x}^T \mathbf{U}^{(2)}_s \delta \mathbf{x} / 2, 
\end{equation}
where $\delta \mathbf{x}$ is the displacement from the saddle point and $\mathbf{U}^{(2)}_s$ is the second derivative matrix of the potential at the saddle point: $[\mathbf{U}^{(2)}_s]_{j,j'}:=\partial ^2 U(\mathbf{x})/\partial x_j \partial x_{j'} |_{\mathbf{x}=\mathbf{x}_s}$. Like in one-dimensional problems, the replacement of the full potential $U(\mathbf{x})$ with  eq.~\ref{U(2)s} is justified in the asymptotic limit $\Delta U \to \infty$ if we scale the potential according to $\Delta U$, that is if we consider potentials of the form
\begin{equation}\label{m-scaling}
    U(\mathbf{x}) =  \Delta U \ s(\mathbf{x})
\end{equation}
for increasing barriers $\Delta U$ but with a fixed potential shape $s(\mathbf{x})$. A simple picture of the diffusion dynamics near the saddle point is recovered from the normal modes defined according to the following eigenvalue problem: 
\begin{equation}\label{normal modes}
    \mathbf{D} \ \mathbf{U}^{(2)}_s \ \mathbf{u}_j = \xi _j \mathbf{u}_j,
\end{equation}
with only one negative eigenvalue $\xi _1$ for the unstable (reactive) mode. The eigenvectors of the non-symmetric  matrix $ \mathbf{D} \mathbf{U}^{(2)}_s$  supply a bi-orthonormal basis as $\mathbf{u}_j^T\mathbf{u}^{j'}=\delta _{j,j'}$, with $\mathbf{u}^{j}= \mathbf{D}^{-1}\mathbf{u}_j$ defining  the displacements $z_j:=\delta \mathbf{x}^T \mathbf{u}^{j}$ along the normal modes. With such a coordinate representation, $\delta \mathbf{x}=\sum _j z_j \mathbf{u}_j$, the FPS operator eq.~\ref{mGamma} in the asymptotic limit is decomposed into independent contributions for each normal mode.  In the Kramers-Langer (KL) procedure, this allows the one-dimensional reduction of the asymptotic kinetic eigenmode $g^\infty_{KL}$  depending only on  the reaction coordinate $z_1$:
\begin{equation}\label{ginftyKL}
    \frac{\partial }{\partial z_1} e^{|\xi _1| z_1^2 /2}  \frac{\partial }{\partial z_1} g^\infty_{KL}=0.
\end{equation}
The solution of this equation has the same form of eq.~\ref{erf},
\begin{equation}\label{merf}
    g^\infty_{KL}=\text{erf} (x^\infty_{KL}), 
\end{equation}
with the scaled reaction coordinate given as $x^\infty_{KL}:=z_1 \sqrt{|\xi_1|/2}$. Finally, from the expectation value of FPS evolution operator (see Section D in Supporting Material),  the Kramers-Langer Asymptotic Relation (KLAR) is derived for the rate constant
\begin{equation}\label{KLAR}
    k^\infty _{KL}=D_{KL} \frac{ |U_{KL}^{(2)}|}{2\pi}  \sqrt{
    {\det (\mathbf{U}^{(2)}_0)}/{|\det (\mathbf{U}^{(2)}_s)|} } \ e^{-\Delta U}, 
\end{equation}
where $D_{KL}$ and $U_{KL}^{(2)}$ are the diffusion coefficient and the saddle point curvature, respectively, 
\begin{equation}\label{DKL}
    \begin{split}
    &1/D_{KL}:=(\mathbf{v}_{KL})^T\mathbf{D}^{-1}\mathbf{v}_{KL}, \\  &U_{KL}^{(2)}:=(\mathbf{v}_{KL})^T \mathbf{U}^{(2)}_s \mathbf{v}_{KL},
    \end{split}
\end{equation}
along the normalized direction, $(\mathbf{v}_{KL})^T  \mathbf{v}_{KL}=1$, of the reaction mode $\mathbf{v}_{KL}\propto \mathbf{u}_1$ of the Kramers-Langer theory. Notice that, taking into account that $\mathbf{D}^{-1}$ is proportional to the friction matrix,  $D_{KL}$ is inversely proportional to the friction along the reaction mode. 

In order to analyze the rate constant of mTGD model, we have first to introduce the asymptotic limit adequate for the multi-dimensional Two Gaussian Distribution. Given the structure eq.~\ref{mTGD} of the distribution, this limit is defined by considering a homogeneous narrowing of the gaussians as obtained by replacing the second moments $\mathbf{\Sigma}$ by $\epsilon \mathbf{\Sigma}$ with a vanishing parameter $\epsilon$. This scaling of the second moments
\begin{equation}\label{Scaling mTGD}
    \mathbf{\Sigma} \ \to \ \epsilon \mathbf{\Sigma} 
\end{equation}
for $\epsilon \to 0^+$, produces a concentration of the two gaussians about their centers which correspondingly reduces their superposition at the saddle point so leading to an increasing barrier as $1/\epsilon$ according to eq.~\ref{barrier mTGD}. It should be stressed that the scaling of the second moments by a scalar parameter does not modify the anisotropy of matrix $\mathbf{\Sigma}$. Of course, parameter $\epsilon$ is instrumental in order to recognize the asymptotic form of the rate constant for increasing barrier and, once it has been found, the original problem is restored by using a unitary $\epsilon$. 

In the next step, we recognize the suitable coordinates for the asymptotic limit. Like with the one-dimensional problem previously analyzed, the odd component $U_-$ of the potential has a critical role in determining the asymptotic form of the kinetic eigenmode. According to eq.~\ref{mTGD potential 3}, $U_-$ includes two kinds of contributions: the first which is linear on the odd coordinates $\delta \mathbf{x}^-$, and the other one given as bilinear products of even $\delta \mathbf{x}^+$ and odd $\delta \mathbf{x}^-$ coordinates. Only the first contribution survives in the one-dimensional problem and, because of its structure of a linear combination, it determines a particular coordinate which can be taken as the reaction coordinate of mTGD problems. To formalize it, we introduce an orthonormal set of vectors $\mathbf{v}_1, \mathbf{v}_2,\cdots, \mathbf{v}_{N^-}$, with $(\mathbf{v}_j)^T\mathbf{v}_{j'}= \delta _{j,j'}$, for the space of odd coordinates, with the following choice of the first vector according to the coefficients of the linear combinations of $\delta \mathbf{x}^-$ in $U_-$
\begin{equation}\label{v_1}
    \mathbf{v}_1=  \frac{1}{a}\left(\mathbf{\Sigma}^{--}\right)^{-1}  \mathbf{x}_0^-,
\end{equation}
where $a:= \sqrt{ (\mathbf{x}_0^-)^T (\mathbf{\Sigma}^{--})^{-2} \mathbf{x}_0^-}$ to ensure the normalization. 
Correspondingly we introduce a different representation of odd coordinates as displacements along vectors $\mathbf{v}_j$
\begin{equation}\label{y_j}
    y_j:= \mathbf{v}_j^T \delta \mathbf{x}^-.
\end{equation}
Then the linear contribution in $U_-$ becomes proportional to $y_1$  and such a coordinate can be employed as the reaction coordinate of the problem, vanishing at the saddle point and taking opposite values by acting on the coordinates with symmetry operator $\hat R$. The potential components, after the scaling eq.~\ref{Scaling mTGD}, become
\begin{equation}\label{mTGD U y}
 \begin{split}
       & U_+ = U(\mathbf{x}_s) +
       \frac{1}{2\epsilon} \left( \delta  \mathbf{x}^+\right)^T \left(\mathbf{\Sigma} ^{-1} \right)^{++} 
        \delta  \mathbf{x} ^+  + \\
       & \ \ \ \ \ \ \ \ \ \ \ \  
       + \frac{1}{2\epsilon} \sum _{j,j'=1}^{N^-}  
       \mathbf{v}_j^T
       \left(\mathbf{\Sigma} ^{-1} \right)^{--} 
       \mathbf{v}_{j'}  y_j y_{j'}, \\
       & U_- =  \frac{a}{\epsilon} y_1 - \frac{1}{\epsilon} 
       \sum _{j=1}^{N^-} \mathbf{v}_j^T
       \left(\mathbf{\Sigma} ^{-1} \right)^{-+} 
        \delta  \mathbf{x}^+ y_j.
\end{split}
\end{equation}
In order to recognize the asymptotic form of the potential, the following scaling of the coordinates according to parameter $\epsilon$ has to be employed
\begin{equation}\label{mTGD scaling}
 \begin{split}
       & y_1^\infty := y_1/\epsilon, \\ 
       & y_j^\infty := y_j/\sqrt{\epsilon} \ \ \ \ 
       \text{for}  \ j=2,3,\cdots,N^-,\\ 
       & { \delta  \mathbf{x}} ^{+ \infty}:= \delta \mathbf{x}^+/\sqrt{\epsilon}.  
\end{split}
\end{equation}
Then by imposing the limit $\epsilon \to 0^+$ to the potential components of eq.~\ref{mTGD U y} after the change to variables eq.~\ref{mTGD scaling}, one obtains their asymptotic forms: 
\begin{equation}\label{mTGD U infty}
 \begin{split}
       & U_+^\infty= U(\mathbf{x}_s) +
        \left( \delta {\mathbf{x}} ^{+ \infty}\right)^T
       \left(\mathbf{\Sigma} ^{-1} \right)^{++}
       \delta {\mathbf{x}} ^{+ \infty} + \\
       & \ \ \ \ \ \ \ \ \ \ \ \  
       + \frac{1}{2} \sum _{j,j'=2}^{N^-}  
       \mathbf{v}_j^T
       \left(\mathbf{\Sigma} ^{-1} \right)^{--} 
       \mathbf{v}_{j'}  y_j^\infty y_{j'}^\infty, \\
       & U_-^\infty= a  y_1^\infty - 
       \sum _{j=2}^{N^-} \mathbf{v}_j^T
       \left(\mathbf{\Sigma} ^{-1} \right)^{-+} 
       \delta  \mathbf{x}^{+ \infty}  y_j^\infty.
\end{split}
\end{equation}
Notice that the direct scaling of the reaction coordinate as $y_1/\epsilon$ is dictated by its linear term contributing to $U_-$ in eq.~\ref{mTGD U y}. On the other hand the different scaling by $\sqrt \epsilon$ for the other coordinates  $y_2,y_3,\cdots,y_{N^-}$ and $\delta  \mathbf{x}^+$ is imposed by the need to recover an equilibrium distribution that can be integrated on these variables. This is the case of the potential components eq.~\ref{mTGD U infty} because of the contribution  by $\exp (-U_+^\infty)$, since $U_+^\infty$ is a positive definite bi-linear form of these coordinates. 

The final step is the calculation of the asymptotic kinetic eigenfunction $g^\infty _{MTGD}$ for the mTGD model. After substitution in eq.~\ref{mGamma} 
of the original $\mathbf{x}$ coordinates with the scaled variables of eq.~\ref{mTGD scaling}, the leading term in the limit $\epsilon \to 0^+$ should be retained. In this way, only the contribution with the derivative with respect to $y_1^\infty$ survives
\begin{equation}\label{mTGD potential 7}
    \frac{\partial}{\partial y_1^\infty}  \cosh (U_-^\infty)
    \frac{\partial}{\partial y_1^\infty} g^\infty_{mTGD} =0
\end{equation}
with $U_-^\infty$ specified by eq.~\ref{mTGD U infty} (see Section D of Supporting Materials). Since this is a differential equation on the variable $y_1^\infty$ only, the dependence of $U_-^\infty$ on the other variables should be treated like for constants of integration and, therefore, the same functional form eq.~\ref{g TGD} of the one-dimensional TGD problem is recovered by imposing the boundary conditions $\lim _{y_1^\infty \to \pm \infty} g^\infty _{mTGD}=\pm 1$:
\begin{equation}\label{g mTGD}
  g^\infty_{mTGD}=\frac{2}{\pi} \int _0 ^{ x^\infty_{mTGD}} \frac{dy}{\cosh(y)},
\end{equation}
where $x^\infty_{mTGD}\equiv U_-^\infty$ is the scaled variable for the asymptotic mTGD kinetic eigenmode, which is linearly dependent on the reaction coordinate $y_1^\infty $ but bears also the dependence on the other  coordinates:
\begin{equation}\label{xmTGD}
  x^\infty_{mTGD} = a y_1^\infty - 
  \sum _{j=2}^{N^-} \mathbf{v}_j^T
       \left(\mathbf{\Sigma} ^{-1} \right)^{-+} 
       \delta  \mathbf{x}^{+ \infty}  y_j^\infty.
\end{equation}
By evaluating in the asymptotic limit $\epsilon \to 0^+$  the expectation value of FPS operator  eq.~\ref{mGamma}  with the previous result for the kinetic eigenmode (see for details Section D of Supporting Materials), the asymptotic rate constant of mTGD model is found
\begin{equation}\label{kmTGD}
 \begin{split}
  k & ^\infty _{mTGD}  = 
   D_{mTGD} 
  \sqrt { \frac 
  {\left(\mathbf{x}_0^-\right)^T \left(\mathbf{\Sigma}^{--}\right)^{-2} \mathbf{x}_0^- /(2\pi^3) } 
  {\det (\mathbf{\Sigma} )  \det(\mathbf{B}) \det\left[ \left({\mathbf{\Sigma}}^{-1}\right)^{++}\right] } }\ e^{-\Delta U}, 
\end{split}
\end{equation}
where we have restored the original mTGD model by attributing a unitary value to parameter $\epsilon$.
In the previous equation $D_{mTGD}$ is the diffusion coefficient along the reaction coordinate $y_1$
\begin{equation}\label{DmTGD}
  D_{mTGD}:=\mathbf{v}_1^T \mathbf{D}\mathbf{v}_1,
\end{equation}
while matrix $\mathbf{B}$ of dimension $(N^--1)\times (N^--1)$ has elements
\begin{equation}\label{B matrix }
  B_{j,j'}:=\mathbf{v}_j^T \left(\mathbf{\Sigma}^{-1}\right)^{++} \mathbf{v}_{j'} 
\end{equation}
for $j,j'=2,3,\dots N^-$.

\subsection{Comparison with the exact numerical rate constants}
\begin{figure}[ht!]
    \centering
    \includegraphics[width=0.95\columnwidth]{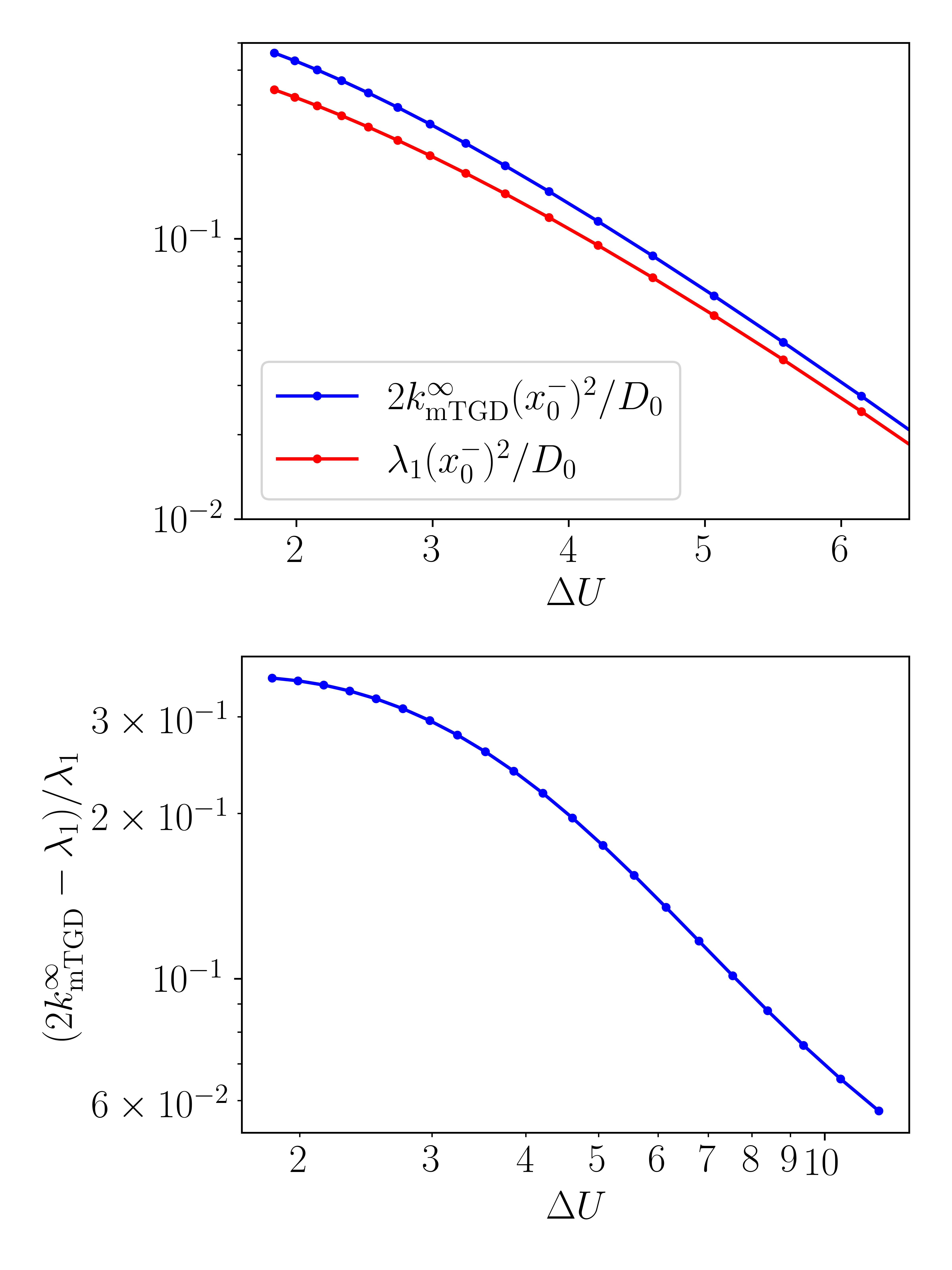}
    \caption{
    Comparison between the kinetic eigenvalue $\lambda _1$ of FPS operator with its asymptotic counterpart $2k^\infty _{mTGD}$ for a two-dimensional realization of mTGD model. The calculations have been done for an isotropic diffusion matrix, $D^{++}=D^{--}=D_0$, fixed angle $\theta=\pi/6$ and first eigenvalue $\sigma_1^2=(x_0^-/5)^2$ of the covariance matrix $\mathbf{\Sigma}$ with a variable second eigenvalue $\sigma_2^2$ to produce an increasing potential barrier height $\Delta U$. Upper panel: comparison of $\lambda _1$ with  $2k^\infty _{mTGD}$ scaled by $D_0/(x_0^-)^2$ as a function of the barrier height $\Delta U$. Lower panel: relative difference $(2k^\infty _{mTGD}-\lambda _1)/\lambda _1$ as a function of the barrier height $\Delta U$ in a logarithmic scale.
    }
    \label{fig: Asymptotic_behavior_TGD_nD_Asy}
\end{figure}

In order to characterize the convergence of the kinetic eigenvalue $\lambda _1$ of the FPS operator to the asymptotic form $2k^\infty _{mTGD}$, we have considered two-dimensional realizations of mTGD model on coordinates $\mathbf{x}=(x^+,x^-)$ like in Fig. 4. More specifically, in order to calculate the exact kinetic eigenvalue $\lambda _1$, we have solved numerically the FPS equation  with an isotropic $2\times 2$ diffusion matrix eq.~\ref{D}, that is for $D^{++}=D^{--}=D_0$, and with a $2\times 2$ covariance matrix $\Sigma$ with fixed values of the angle, $\theta=\pi/6$, and of the first eigenvalue, $(\sigma_1/x_0^-)^2=0.04$, for decreasing values of the second eigenvalue $(\sigma_2/x_0^-)^2$ in the range $[0.04,0.25]$ in order to reproduce an increasing barrier $\Delta U$ evaluated according to eq.~\ref{barrier mTGD}. In the upper panel of Fig.~\ref{fig: Asymptotic_behavior_TGD_nD_Asy}, we have represented both the kinetic eigenvalue $\lambda _1$ of FPS operator (red dots) and its asymptotic counterpart $2k^\infty_{mTGD}$ of eq.~\ref{kmTGD} (blue dots), both scaled by $D_0/(x_0^-)^2$, as a function of the barrier height $\Delta U$. Clearly, by increasing the barrier height, the kinetic eigenvalue $\lambda _1$ gets closer to its asymptotic estimate $2k^\infty_{mTGD}$. The convergence of the kinetic eigenvalue to its asymptotic limit  is more directly verified in the lower panel of Fig. 5 where the relative difference $(2k^\infty_{mTGD}-\lambda _1)/\lambda _1$ is represented for increasing barriers in a logarithmic scale.

The results reported in Fig.~\ref{fig: Asymptotic_behavior_TGD_nD_Asy} show that our analysis of mTGD models supplies the correct asymptotic form of the rate constant. On the other hand, one might apply the Kramers-Langer asymptotic rate constant eq.~\ref{KLAR} to mTGD models by simply evaluating the curvature matrices according to eq.~\ref{mTGD potential 2}, but an Arrhenius form with a different pre-exponential factor would result. Indeed, the Kramers-Langer method and our analysis of mTGD model are intrinsically different procedures just because they are based on different definitions of the asymptotic limit: the scaling eq.~\ref{m-scaling} of the potential by the barrier height in the former and the scaling eq.~\ref{Scaling mTGD} of the second moments of the gaussian distributions in the latter. In both cases, however, the kinetic eigenmode in the asymptotic limit is described by one-dimensional functional forms even if of different nature, eq.~\ref{merf} and eq.~\ref{g mTGD} in the two cases. The identification of the reaction coordinate exemplifies the difference of these two procedures. In the Kramers-Langer method the reaction coordinate is determined by the unstable normal mode at the saddle point, so depending on both the diffusion matrix and the curvature matrix according to eq.~\ref{normal modes}, while in the case of mTGD model it is independent of the diffusion matrix since it is derived from the asymptotic limit of the potential, which  leads to  eq.~\ref{v_1} for the direction of the reaction coordinate. 
The reaction coordinates resulting from the two methods, however, are the same in the particular case of mTGD models with one odd coordinate only ($N^-=1$) because  both the diffusion matrix $\mathbf{D}$  and the curvature matrix $\mathbf{U}_s^{(2)}$ at the saddle point, to be employed for the determination of the unstable normal model, are block separated with respect to even/odd coordinates. Still different results for the asymptotic rate constant are recovered from the two methods, very much like the one-dimensional case that has been initially considered.

\section{Conclusions}\label{sec: Conclusions}

A large part of the literature about the stochastic analysis of activated processes is based on models of the mean-field potential specified through suitable parameterized forms~\cite{Hanggi1990}. In this work, we follow the alternative route of modeling directly the equilibrium distribution through the linear combination of two gaussians parameterized according to their second moments providing their widths. Even if the link between equilibrium distribution and mean-field potential is clearly established by the Boltzmann relation, the two procedures are not equivalent from the point of view of the parametric space of the models. In particular, the linear scaling of the mean-field potential is not in general allowed by parameterized forms of the equilibrium distribution. By using the stochastic description provided by the Fokker-Planck-Smoluchowski equation, we have characterized the diffusion models deriving from the parameterized Two Gaussian
Distribution, first in one-dimensional problems (the TGD model) and afterwords in multidimensional problems (the m-TGD model). 
The definition of the stochastic model through the direct parameterization of the equilibrium distribution becomes necessary when a physical model of the corresponding potential is not available as in the case of quantum tunneling~\cite{Pravatto2022}. However, we have shown here that this model also provides a flexible and convenient tool for studying the rate of activated processes as an alternative to the more conventional parametrized forms of mean field potential. In particular, the parameterization of the equilibrium density in terms of a linear combination of Gaussian functions allows a straightforward generalization to multidimensional problems and facilitates the evaluation of expectation values.

A peculiarity of this kind of modeling activated processes through the equilibrium distribution is that the large barrier limit is not reproduced by the Kramers (KAR) and the Kramers-Lamger (KLAR) Asymptotic Relations. The one-dimensional case of TGD is particularly illuminating since in practice it has only one control parameter which can be identified with the barrier height. The comparison with the exact numerical values of the rate constant provides  clear evidence that the Kramers asymptotic result (KAR) does not provide the right pre-exponential factor of the Arrhenius form. The origin of the discrepancy resides on the fact that the Kramers and the Kramers-Langer asymptotic analysis are justified by the linear scaling of the mean-field potential, a procedure that is not allowed in the case of TGD and m-TGD models. For this type of model, one needs a different asymptotic procedure which we have presented in this contribution together with its validation by comparison with the numerical results for one- and two-dimensional realizations of the proposed model.

Asymptotic results, even if they provide only approximations to the rate constants in the applications of stochastic theories as long as one is dealing always with finite barriers, are important tools for the identification of the features of the model which control the kinetic process.

\section*{Acknowledgments}
This work was supported by the Department of Chemical Sciences (DiSC) and the University of Padova through the DOR funding scheme and Project QA-CHEM (P-DiSC\#04BIRD2021-UNIPD). P. Pravatto is grateful to Fondazione CARIPARO for the financial support (PhD grant) and to the Cloud Veneto computing infrastructure for the computational time granted under the project "Nuclear Quantum Effects".

\printbibliography
\newpage

\section*{Supporting Material}

{\bf Section A: Schroedinger representation of the diffusion equation.}

For the formal analysis of the Fokker-Planck-Smoluchowski (FPS) eq.~\ref{FPS}, it is convenient to deal with its symmetrized form which generates the Schroedinger representation of FPS ~\cite{Risken1989, Elber2020}. Let us introduce the modified form $\tilde \rho_t(x)$ of the probability density $\rho _(x)$, defined  as 
\begin{equation}\label{SMA1}
    \tilde \rho_t(x) :=  \rho_{eq}(x)^{-1/2}  \rho_t(x).
\end{equation}
Its time evolution 
\begin{equation}\label{SMA2}
    \frac{\partial}{\partial t} \tilde{\rho} _t (x) = - \hat{\tilde{\Gamma}} \tilde{\rho}_t (x)
\end{equation}
is driven by the following symmetrized  Fokker-Planck-Smoluchowski operator
\begin{equation}\label{SMA3}
     \begin{split}
    \hat{\tilde{\Gamma}} & = \rho_{eq}(x)^{-1/2} \hat{\Gamma} \rho_{eq}(x)^{1/2} = \\ 
    &=- \rho_{eq}(x)^{-1/2} \frac{\partial}{\partial x} D \rho_{eq}(x) \frac{\partial}{\partial x} \rho_{eq}(x)^{-1/2},
    \end{split}
\end{equation}
where the ordinary FPS operator $\hat \Gamma$ is specified according to eq.~\ref{Gamma}. Such an operator $\hat{\tilde{\Gamma}}$ is self-adjoint (hermitian) with respect to the scalar product defined as 
\begin{equation}\label{SMA4}
    \langle f \rvert g \rangle:= \int ^{+\infty} _{-\infty} ds f(x)^* g(x)
\end{equation}
with the elements $f(x)$ and $g(x)$ of the Hilbert space vanishing at infinity ($x\to \pm \infty$) to ensure integrability. By specifying the equilibrium distribution according to the mean field potential, $\rho_{eq}(x) \propto \exp \{-U(x)\}$, the symmetrized operator can be rearranged as 
\begin{equation}\label{SMA5}
    \hat{\tilde{\Gamma}}  = -D \frac{\partial^2}{\partial x^2} +\frac{D}{4} U'(x)^2 - \frac{D}{2}U''(x)
\end{equation}
which has the form of typical one-dimensional Hamiltonians of Quantum Mechanics, provided that the diffusion coefficient is rewritten as $D=\hbar^2/2m$  so that $-D\partial ^2/\partial x^2$ corresponds to the kinetic energy contribution with the remaining part at the r.h.s identifying the quantum potential of the problem. Such an isomorphism between the FPS time evolution operator and the Hamiltonian operator was exploited in ref.~\cite{Pravatto2022} to evaluate a quantum property like tunneling splitting by employing stochastic tools for activated processes. 

The isomorphism with a Hamiltonian operator can be used to recognize the independent eigenmodes of FPS in eq.~\ref{eiegenfunctions}
as the solutions of time-independent Schroedinger-like equation 
\begin{equation}\label{SMA6}
    \hat{\tilde{\Gamma}} \rho _{eq}(x)^{1/2} \phi _n(x) =\lambda _n \rho _{eq}(x)^{1/2} \phi _n(x), 
\end{equation}
where the equilibrium distribution weight $\rho _{eq}(x)^{1/2}$ has been singled out from the eigenfunctions. Since operator $\hat{\tilde{\Gamma}} $ is hermitian, the ensemble of its eigenfunctions represents an orthonormal basis
\begin{equation}\label{SMA7}
    \langle \rho _{eq}^{1/2}\phi _n \rvert \rho _{eq}^{1/2} \phi _{n'} \rangle =\delta_{n,n'},
\end{equation}
while its eigenvalues result to be positive semi-definite
\begin{equation}\label{SMA8}
      \lambda _n= \langle \rho _{eq}^{1/2}\phi _n \rvert \hat{\tilde{\Gamma}} \lvert \rho _{eq}^{1/2} \phi _n \rangle= D \langle \phi _n' \rvert  \rho _{eq} \lvert \phi'_n\rangle \ge 0,
\end{equation}
where $\phi _n'(x):=\text{d}\phi _n(x)/\text{d}x$  and integration by parts has been employed to derive the last equation. Of course, the inequality is due to the fact that $\langle \phi _n'\rvert \rho _{eq}\lvert \phi _n'\rangle$ is given as the integral of a non-negative function. We recall also that the trivial solution 
\begin{equation}\label{SMA9}
\phi _0 (x)=1, \qquad \lambda _0=0
\end{equation}
corresponding to the quantum ground state is recovered for the stationary solution of FPS equation according to the equilibrium distribution.

Then the time dependence of a generic probability density $\rho_t(x)$ can be made explicit by decomposing it on the basis of the eigenfunction,
\begin{equation}\label{SMA10}
    \begin{split}
    \rho & _t(x)=\rho_{eq}(x)^{1/2}\tilde \rho_t(x) = \\ 
    &= \rho_{eq}(x)^{1/2} \sum _{n=0}^\infty \rho_{eq}(x)^{1/2}\phi _n(x) e^{-\lambda _n t} \langle \rho_{eq}^{1/2}\phi _n \rvert \tilde \rho _0 \rangle = \\
    &=\rho_{eq}(x)\sum _{n=0}^\infty \phi _n(x) e^{-\lambda _n t} \langle \phi _n\rvert \rho _0\rangle,     \end{split}  
\end{equation}
where $\rho _0(x)=\rho _t(x)|_{t=0}$. So a multiexponential evolution is derived in all generality but, in the presence of the time scale separation $\tau _{lr}:=1/\lambda _2>>\tau _{kin}:=1/\lambda _1$, for times much longer than the local relaxation, $t>>\tau _{lr}$, a single exponential decay characteristic of the kinetic regime is recovered according to eq.~\ref{regime cinetico}. This allows one to establish a  direct relation eq.~\ref{k} between the rate constant $k$ and the first non-vanishing eigenvalue $\lambda _1$.

Finally, we mention that given an approximation $\rho_{eq}(x)^{1/2}\eta (x)$ of the first excited state $\rho_{eq}(x)^{1/2}\phi _1(x)$, under the condition of orthogonality $\langle \rho_{eq}^{1/2}\eta \rvert \rho_{eq}^{1/2}\phi\rangle=0$ with respect to the ground state $\rho_{eq}(x)^{1/2}\phi _0(x)$, for instance, because of its odd parity $\eta (-x)=-\eta (x)$, then according to the Rayleigh-Ritz theorem the corresponding expectation value  supplies a majorant of the first non-vanishing eigenvalue:
\begin{equation}\label{SMA11}
\lambda _1 \le \frac{\langle \rho_{eq}^{1/2}\eta \rvert  \hat{\tilde{\Gamma}} \lvert \rho_{eq}^{1/2}\eta \rangle}{\langle \rho_{eq}^{1/2}\eta \rvert \rho_{eq}^{1/2}\eta \rangle}= D \frac{\langle \eta ' \rvert \rho_{eq} \lvert\eta '\rangle }{\langle \rho_{eq}^{1/2}\eta \rvert\rho_{eq}^{1/2}\eta\rangle}.
\end{equation}
Therefore, if an approximation of the kinetic component $\phi_1(x)$ is available, the calculation of its expectation value like in the r.h.s. of eq.~\ref{SMA11} would supply an estimate of $\lambda _1$, and of the rate constant as well according to  eq.~\ref{k}, with strictly positive deviations.\\

{\bf Section B: Asymptotic rate constant of one-dimensional diffusion.}

We evaluate the asymptotic rate constant according to the expectation value of FPS operator from asymptotic forms $g^\infty(x) =g^\infty_K(x)$ of eq.~\ref{erf} and  $g^\infty(x) =g^\infty_{TGD}(x)$ of eq.~\ref{g TGD} for the kinetic eigenmode
\begin{equation}\label{SMB1}
    \begin{split}
    2k^\infty&=\langle g^\infty \rvert \hat \Gamma \rvert \rho_{eq} g^\infty \rangle / \langle g^\infty \rvert \rho_{eq} g^\infty \rangle = \\
    &= D \langle {g^\infty}' \rvert  \rho_{eq} {g^\infty}' \rangle / \langle g^\infty \rvert \rho_{eq} g^\infty \rangle,
    \end{split}
\end{equation}
where ${g^\infty}'(x) =dg^\infty(x)/dx$ and to obtain the r.h.s. we have performed an integration by parts after inserting operator $\hat \Gamma $ of eq.~\ref{Gamma}. In the asymptotic limit the denominator is unitary since $g^\infty (x)\not= 1$ only in a small and decreasing range about the saddle point and, therefore
\begin{equation}\label{SMB2}
    k^\infty = \frac{D}{2} \int _{-\infty}^{+\infty}\ dx \frac{e^{-U(x)}}{Z} {g^\infty}'(x)^2, 
\end{equation}
where $Z=\int _{-\infty}^{+\infty} dx \exp [-U(x)]$ is the normalization of equilibrium distribution eq.~\ref{rhoeq}.

Let us consider the result of the Kramers procedure by inserting into the previous equation the corresponding asymptotic form $g^\infty(x) =g^\infty_K(x)$  eq.~\ref{erf}  of the kinetic eigenmode
\begin{equation}\label{SMB3}
    k^\infty _K= D \sqrt {2 \left| U_s^{(2)}\right |\Big{/} \pi } \ \frac{e^{-U(0)}}{Z},
\end{equation}
where the integration has been done by employing the parabolic form eq.~\ref{U parabolic} of the potential. 
On the other hand the normalization factor $Z$ can be evaluated from the parabolic expansion $U(x)=U(x_m)+U_m^{(2)}(x\pm x_m)^2/2$ of the potential about the two minima at $\pm x_m$
\begin{equation}\label{SMB4}
    Z=  \sqrt {8 \pi \Big{/}  U_m^{(2)} } \ e^{-U(x_m)}.
\end{equation}
Finally, by inserting it into eq.~\ref{SMB3} and by taking into account that the barrier height is defined as $\Delta U := U(0)-U(x_m)$, the Asymptotic Kramers Relation (KAR) eq.~\ref{KAR} is obtained
\begin{equation}\label{SMB5}
    k^\infty _K= \frac{D}{2\pi} \sqrt{U_0^{(2)}\left| U_s^{(2)} \right| }\  e^{-\Delta U}.
\end{equation}

Let us now specify eq.~\ref{SMB2} for the TGD potential whose curvatures at the saddle point and at the minima for $\sigma/x_0<<1$ are given as $U_s^{(2)}=-x_0^2/\sigma^4$  and $U_m^{(2)}=1/\sigma^2$, respectively. By replacing $g^\infty (x)$ in eq.~\ref{SMB2}  with $g^\infty _{TGD} (x)$ of eq.~\ref{g TGD}, the asymptotic rate constant for TGD model is obtained
\begin{equation}\label{SMB6}
    k^\infty _{TGD}= D \frac{2 x_0}{\pi \sigma ^2} \ \frac{e^{-U(0)}}{Z}
\end{equation}
which, after substitution of the normalization factor of  eq.~\ref{SMB4} with the proper potential curvature of TGD potential, leads to 
eq.~\ref{k TGD}, that is 
\begin{equation}\label{SMB7}
  k^\infty _{TGD}=D\frac{x_0}{ \sqrt{2\pi ^3}\ \sigma ^3} \ e^{-\Delta U}.
\end{equation}
Notice that by inserting into KAR eq.~\ref{SMB5} the curvatures of TGD potential, the following asymptotic rate constant would be obtained
\begin{equation}\label{SMB8}
  k^\infty _{K}=D\frac{x_0}{2 \pi \sigma ^3 } \ e^{-\Delta U},
\end{equation}
which overestimates the correct asymptotic result of eq.~\ref{SMB7} by a factor $\sqrt{\pi/2}$.\\

{\bf Section C: Mean field potential and barrier height of the multi-dimensional TGD (mTGD) model.}

By specifying into the definition  eq.~\ref{mTGD potential} of the potential the equilibrium distribution of eq.~\ref{mTGD} and by separating in the bi-linear forms at the exponents the contributions of the coordinates $\mathbf{x}$ from those of the first moments $\mathbf{x}_0$, as well as the contributions of even and odd components, one obtains the following relation
\begin{equation}\label{SMC1}
    U(\mathbf{x})=U_+(\mathbf{x})-\ln \{\cosh [U_-(\mathbf{x})] \},
\end{equation}
with 
\begin{equation}\label{SMC2}
    \begin{split}
    U_+&(\mathbf{x})  =U(0) +
    \left( \mathbf{x}^+ \right )^T \left( \mathbf{\Sigma}^{-1} \right) ^{++} \mathbf{x}^+/2 + \\
    & + \left( \mathbf{x}^- \right)^T \left( \mathbf{\Sigma}^{-1} \right)^{--} \mathbf{x}^-/2 -  
    \left(\mathbf{x}^+\right)^T \left(\mathbf{\Sigma}^{-1}\right)^{+-} \mathbf{x}_0^-
    \end{split}
\end{equation}
and
\begin{equation}\label{SMC3}
    U_- (\mathbf{x})  =-
    \left(\mathbf{x}^+\right)^T \left(\mathbf{\Sigma}^{-1}\right)^{+-} \mathbf{x}^-
    + \left(\mathbf{x}^-\right)^T \left(\mathbf{\Sigma}^{-1}\right)^{--} \mathbf{x}^-_0,
\end{equation}
where $U(0)$ is the potential at the origin of coordinates, that is for $\mathbf{x}=0$,
\begin{equation}\label{SMC4}
    U(0)=(\mathbf{x}_0)^T \mathbf{\Sigma}^{-1} \mathbf{x}_0/2 - \ln 2.
\end{equation}
Functions $U_+(\mathbf{x})$ and $U_-(\mathbf{x})$ can be interpreted as the even and odd, respectively, components of the potential, since $\hat R U_\pm(\mathbf{x})=\pm U_\pm(\mathbf{x})$.

The saddle point  has necessarily a vanishing odd component, $\mathbf{x}_s=(\mathbf{x}_s^+,0)$, since otherwise the symmetry of the potential, $\hat R U(\mathbf{x})=U(\mathbf{x})$, would imply the existence of two degenerate saddle points. Therefore, in order to find it, one can consider only the gradient with respect to even coordinates at $\mathbf{x}^-=0$:
\begin{equation}\label{SMC5}
    \begin{split}
    \frac{\partial U(\mathbf{x})}{\partial \mathbf{x}^+}   \bigg|_{\mathbf{x}^-=0} &=
    \frac{\partial U_+(\mathbf{x})}{\partial \mathbf{x}^+}  \bigg| _{\mathbf{x}^-=0} = \\
    &= (\mathbf{\Sigma}^{-1})^{++} \mathbf{x}^+ -  (\mathbf{\Sigma}^{-1})^{+-} \mathbf{x}^-_0. 
    \end{split}
\end{equation}
Then the condition of the vanishing gradient leads to 
eq.~\ref{saddle} for the location of the saddle point.

The calculation of the potential barrier
\begin{equation}\label{SMC6}
    \Delta U := U(\mathbf{x}_s) - U(\mathbf{x}_0)
\end{equation}
requires the evaluation of the potential at the saddle point $\mathbf{x}_s$ and at $\mathbf{x}_0$.
Let us first show that $U(\mathbf{x}_0)$ has a negligible value. Indeed, by replacing $\rho _{eq}(\mathbf{x}_0)$ directly into eq.~\ref{mTGD potential} one obtains
\begin{equation}\label{SMC7}
    U(\mathbf{x}_0)  = - \ln (1+S),
\end{equation}
where $S=\exp \{ -2 (\mathbf{x}_0)^T \mathbf{\Sigma}^{-1}\mathbf{x}_0 \}$ is the superposition parameter of the two gaussian, which is supposed to be negligible, so that $ U(\mathbf{x}_0) \simeq 0$. Therefore the potential barrier can be identified with the potential at the saddle point:
\begin{equation}\label{SMC8}
    \Delta U = U(\mathbf{x}_s)=  U_+(\mathbf{x}_s),
\end{equation}
where the r.h.s. takes into account that $ U_-(\mathbf{x}_s)=0$. By evaluating $ U_+(\mathbf{x}_s)$ according to the saddle point specified by eq.~\ref{saddle}, one obtains
\begin{equation}\label{SMC9}
    \Delta U = \frac{1}{2} (\mathbf{x}_0^-)^T  \left [ \mathbf{A}^{--} - \mathbf{A}^{-+}(\mathbf{A}^{++})^{-1}  \mathbf{A}^{+-} \right ] \mathbf{x}_0^-  - \ln (2),
\end{equation}
where $\mathbf{A}:=\mathbf{\Sigma}^{-1} $. Then, by recalling the block matrix inversion, the explicit form eq.~\ref{barrier mTGD} is recovered for the barrier height. 

Finally, we need to specify the dependence of even $U_+$ and odd $U_-$ components of the potential on the displacements from the saddle point
\begin{equation}\label{SMC10}
    \delta \mathbf{x} := \mathbf{x} -\mathbf{x}_s.
\end{equation}
 The first of eq.~\ref{mTGD potential 3} is readily obtained from Eq.~\ref{SMC2} after the substitution $\mathbf{x} = \delta  \mathbf{x} +\mathbf{x}_s$  and by taking into account that $U(\mathbf{x}_s)=U_+(\mathbf{x}_s)$. The same substitution into eq.~\ref{SMC3} leads to the relation
\begin{equation}\label{SMC11}
    \begin{split}
    U_- & =  
     -\left( \delta \mathbf{x}^+\right)^T \mathbf{A}^{+-} \delta \mathbf{x}^- + \\
    & + 
    \left(\delta \mathbf{x}^-\right)^T  \left [ \mathbf{A}^{--} - \mathbf{A}^{-+}\left(\mathbf{A}^{++}\right)^{-1}  \mathbf{A}^{+-} \right ] \mathbf{x}_0^-. 
    \end{split}
\end{equation}
Again by recalling the block matrix inversion like in  eq.~\ref{SMC9}, the final form eq.~\ref{mTGD potential 3} for the odd component of the potential is recovered. \\

{\bf Section D: Asymptotic rate constant for multidimensional diffusion.}\\

In the first part of this Section,  the treatment leading to  Kramers-Langer Asymptotic Relation (KLAR) will be presented.
Let us first reformulate the eigenvalue problem eq.~\ref{normal modes} in matrix notation 
\begin{equation}\label{SMD1}
    \mathbf{D} \ \mathbf{U}^{(2)}_s \ \mathbf{S} =  \mathbf{S} \mathbf{X},
\end{equation}
where $ \mathbf{X}$ is the diagonal matrix with the eigenvalues and $ \mathbf{S}$ matrix collects eigenvectors $\mathbf{u}_j$ by columns. The bi-orthogonality condition 
$\mathbf{u}_j^T\mathbf{u}^{j'}=
\mathbf{u}_j^T\mathbf{D}^{-1}\mathbf{u}_{j'}=
\delta _{j,j'}$ becomes
\begin{equation}\label{SMD2}
    \mathbf{S}^T \mathbf{D}^{-1} \mathbf{S} =\mathbf{1}.
\end{equation}
The ensemble $\mathbf{z}^T=(z_1,z_2,\cdots , z_N)$ of displacements along the normal modes are then specified as 
\begin{equation}\label{SMD3}
    \mathbf{z}= \mathbf{S}^T \mathbf{D}^{-1}\delta \mathbf{x}.
\end{equation}
By replacing in FPS operator eq.~\ref{mGamma} the $\mathbf{x}$ representation with the $\mathbf{z}$ representation under the asymptotic limit when the second order expansion eq.~\ref{U(2)s} holds for the potential, a diffusion operator with independent contributions is recovered,
\begin{equation}\label{SMD4}
     \hat \Gamma= - \sum _j \frac{\partial }{\partial z_j} e^{-|\xi _j| z_j^2 /2}  \frac{\partial }{\partial z_j} e^{|\xi _j| z_j^2 /2}, 
\end{equation}
and this justifies the use of eq.~\ref{ginftyKL} for obtaining the asymptotic kinetic eigenmode $g^\infty _{KL}$ depending on the reaction coordinate $z_1$.

The KLAR eq.~\ref{KLAR} for the rate constant $k^\infty _{KL}$ is then recovered by evaluating in the asymptotic limit the expectation value of $\hat \Gamma$ with the Kramers-Langer form $g^\infty_{KL}$ of the kinetic eigenmode
\begin{equation}\label{SMD5}
     2 k^\infty _{KL}= \langle g^\infty_{KL} \rvert \hat \Gamma \lvert  g^\infty_{KL} \rho _{eq}   \rangle, 
\end{equation}
where $\langle \dots \rangle =\int d\mathbf{x} \cdots$ and the unitary normalization $\langle g^\infty_{KL} \vert  g^\infty_{KL} \rho _{eq}   \rangle =1$ was assumed since, for $\Delta U \to \infty$, $\rvert g^\infty_{KL} \lvert =1 $ away from the saddle point. By using $g^\infty_{KL}$ of eq.~\ref{merf} and by performing integration by parts of FPS operator of eq.~\ref{SMD4}, one obtains
\begin{equation}\label{SMD6}
    \begin{split}
    k^\infty _{KL} & = \frac{1}{2} \langle \partial g^\infty_{KL}/\partial z_1  \rvert  \rho _{eq}
    \lvert \partial g^\infty_{KL}/\partial z_1  \rangle = \\
    &= \frac{|\xi_1|}{\pi} \langle \exp \{-|\xi_1|z_1^2 \} \rho _{eq} \rangle = \\
    &= \frac{|\xi_1|}{\pi}  \rho _{eq}(\mathbf{x}_s) \langle \exp \{-\sum _j|\xi_j|z_j^2/2 \} \rangle,
    \end{split}
\end{equation}
where the r.h.s has been derived by specifying  the equilibrium distribution $\rho _{eq}(\mathbf{x})$ according to the second order expansion eq.~\ref{U(2)s} of the potential. The integration is conveniently performed on the normal modes coordinates $\mathbf{z}$ specified by eq.~\ref{SMD3} 
\begin{equation}\label{SMD7}
    \begin{split}
     \langle \exp & \{-\sum _j|\xi_j|z_j^2/2 \} \rangle = \\
    & = \frac{\det (\mathbf{D})} {|\det (\mathbf{S})|} \int d\mathbf{z} \exp \{-\sum _j|\xi_j|z_j^2/2 \} = \\
    & = \frac{\det (\mathbf{D})} {|\det (\mathbf{S})|} \frac{(2\pi)^{N/2}}{\sqrt {|\det (\mathbf{X})|}}
    =\frac{(2\pi)^{N/2}}{\sqrt {\left|\det \left(\mathbf{U}_s^{(2)}\right)\right|}},
    \end{split}
\end{equation}
where the determinants have been evaluated according to eq.~\ref{SMD1}  and eq.~\ref{SMD2}. On the other hand,  the equilibrium distribution $\rho _{eq}(\mathbf{x}_s) = \exp \{-U((\mathbf{x}_s)\}/Z $ at the saddle point requires the normalization $Z$ which can be determined from the second order expansion of the potential at a minimum
\begin{equation}\label{SMD8}
    \begin{split}
    & U(\mathbf{x})= U(\mathbf{x}_0)+ (\mathbf{x}-\mathbf{x}_0)^T \mathbf{U}_0^{(2)} (\mathbf{x}-\mathbf{x}_0) /2, \\
    & Z=\int d\mathbf{x} \exp \{-U(\mathbf{x}) \} = 2 
    \frac{  (2 \pi)^{N/2} } {\sqrt {\det \left(\mathbf{U}_0^{(2)}\right)}} e^{- U(\mathbf{x}_0)},
    \end{split}
\end{equation}
where factor 2 is due to the degeneracy of the minima at $\pm \mathbf{x}_0$. By substituting these results into eq.~\ref{SMD6}, we get the final result for KLAR
\begin{equation}\label{SMD9}
    k^\infty _{KL}  =  \frac{|\xi_1|}{2\pi} \sqrt{  \frac{\det\left(\mathbf{U}_0^{(2)}\right)}  {\left|\det\left(\mathbf{U}_s^{(2)}\right)\right|}  } \ e^{-\Delta U},
\end{equation}
where $\Delta U = U(\mathbf{x}_s)-U(\mathbf{x}_0)$ is the barrier height. In this form, the rate constant is proportional to the eigenvalue $\xi _1$  of eq.~\ref{normal modes} assigned  to the reaction mode $\mathbf{u}_1$. In order to get more physical insight, it is convenient to specify it according to the direction $\mathbf{v}_{KL}\propto \mathbf{u}_1$ of the reaction mode with Euclidean normalization, $(\mathbf{v}_{KL})^T \mathbf{v}_{KL}=1$. After multiplication by $(\mathbf{v}_{KL})^T\mathbf{D}^{-1}$ of the equation for the reaction mode specified as $\mathbf{D}\mathbf{U}_s^{(2)}\mathbf{v}_{KL}=\xi_1 \mathbf{v}_{KL}$, one obtains
\begin{equation}\label{SMD10}
    \xi _1 = D_{KL} {U}_{KL}^{(2)},
\end{equation}
where ${U}_{KL}^{(2)}$ and $ D_{KL} $ are the curvature and the diffusion coefficient, respectively, along the reactive mode as defined by  eq.~\ref{DKL}. The final substitution of this relation for $\xi _1$ eigenvalue into eq.~\ref{SMD9}, leads to the Kramers-Langer Asymptotic Relation specified by 
eq.~\ref{KLAR} of the main text.

In the following part of this Section, we evaluate the asymptotic rate constant $k^\infty _{mTGD}$ for mTGD model as the expectation value of FPS operator with the kinetic eigenfunction  $g^\infty _{mTGD}$ of eq.~\ref{g mTGD}, like with eq.~\ref{SMD5} for the Kramers-Langer procedure,
\begin{equation}\label{SMD11}
    \begin{split}
     2 k^\infty _{mTGD} &= \langle g^\infty_{mTGD} \rvert \hat \Gamma \lvert  g^\infty_{mTGD} \ \rho _{eq}  \rangle = \\
     &= \int d\mathbf{x} \ \rho _{eq} (\mathbf{x}) \  
     \frac{\partial g^\infty_{mTGD} }{\partial \mathbf{x} } ^T \mathbf{D}  \frac{\partial g^\infty_{mTGD} }{\partial \mathbf{x} },
     \end{split}
\end{equation}
where the r.h.s. has been obtained by integration by parts. We recall that the asymptotic limit is defined by the condition $\epsilon \to 0^+$, with parameter $\epsilon$ scaling the matrix of second moments according to  eq.~\ref{Scaling mTGD}. First of all we perform the change of variables from the original coordinates $\mathbf{x}$ to the $\epsilon$ scaled coordinates of eq.~\ref{mTGD scaling} by retaining in the differential term at the r.h.s. of eq.~\ref{SMD11} the leading contribution for $\epsilon \to 0^+$. By taking into account that the derivative $\partial /\partial y_1^\infty$ brings a factor $1/\epsilon$ while a derivative with respect to the other variables has a factor $1/\sqrt \epsilon$, the leading term is specified as:
\begin{equation}\label{SMD12}
    \begin{split}
     \frac{\partial g^\infty_{mTGD} }{\partial \mathbf{x} } ^T \mathbf{D}  \frac{\partial g^\infty_{mTGD} }{\partial \mathbf{x} } & = 
      \frac{1}{\epsilon ^2} D_{mTGD}\left ( 
      \frac{\partial g^\infty_{mTGD} }{\partial y_1^\infty } \right )^2 \\
      & = \left( \frac{2a}{\pi \epsilon} \right)^2
      \frac{ D_{mTGD} }{\cosh ^2(U_-^\infty) },
     \end{split}
\end{equation}
where $D_{mTGD}$ of eq.~\ref{DmTGD} is the diffusion along the reaction coordinate $y_1$ and parameter $a$ is the normalization introduced in 
eq.~\ref{v_1}. Correspondingly the asymptotic rate constant is specified as
\begin{equation}\label{SMD13}
    \begin{split}
     k^\infty _{mTGD} &=2 \ \epsilon ^{(N+1)/2}  
     \left( \frac{a}{\pi \epsilon} \right)^2
     D_{mTGD} \ \rho _{eq} (\mathbf{x}_s ) \times \\
     & \times \prod _{k=1}^{N^+} 
     \left ( \int _{-\infty} ^{+\infty}  dx_k^{+ \infty} \right )
     \prod _{j=2}^{N^-} 
     \left ( \int _{-\infty}^{+\infty}  dy_j^\infty \right ) \times \\
     & \times \exp (-U^\infty _+ ) \int _{-\infty} ^{+\infty} dy_1^\infty 
     \frac{1}{\cosh (U_-^\infty)},
     \end{split}
\end{equation}
where the equilibrium distribution has been specified as
\begin{equation}\label{SMD14}
    \rho _{eq} (\mathbf{x}) = \rho _{eq} (\mathbf{x}_s) \frac{e^{U(\mathbf{x}_s)-U^\infty _+}}{\cosh (U^\infty _-)},
\end{equation}
with the asymptotic components $U^\infty _+$ and $U^\infty _-$ of the potential given by eq.~\ref{mTGD U infty}. Notice that the factor $\epsilon ^{(N+1)/2}$ is the Jacobian for the change of variables from the original coordinates to those of eq.~\ref{mTGD scaling}. Taking into account that  
\begin{equation}\label{SMD15}
    \begin{split}
    & \int _{-\infty} ^{+\infty} dy_1^\infty 
     \frac{1}{\cosh (U_-^\infty)} = 
     \frac{\pi}{a}, \\
     & \prod _{j=2}^{N^-} 
     \left ( \int _{-\infty}^{+\infty}  dy_j^\infty \right ) \times \\
     & \ \ \ \ \ \ \ \times \exp \left \{ - 
     \frac{1}{2}\sum _{j,j'=2}^{N^-}  
       \mathbf{v}_j^T
       (\mathbf{\Sigma} ^{-1} )^{--} 
       \mathbf{v}_{j'}  y_j^\infty y_{j'}^\infty 
     \right \} = \\
      & \ \ \ \ \ \ \ = 
     \sqrt {(2\pi)^{N^--1} / \det ({\mathbf{B})} }, \\
     & \prod _{k=1}^{N^+} 
     \left ( \int _{-\infty} ^{+\infty}  dx_k^{+ \infty} \right ) \times \\
      & \ \ \ \ \ \ \ \times 
     \exp \left \{ - \delta ({\mathbf{x}} ^{+ \infty})^T
       (\mathbf{\Sigma} ^{-1} )^{++}
       \delta {\mathbf{x}} ^{+ \infty} \right \}  = \\
      & \ \ \ \ \ \ \ =   
     \sqrt {(2\pi)^{N^+} / \det[ ({\mathbf{\Sigma}}^{-1})^{++}]   },
     \end{split}
\end{equation}
where matrix $\mathbf{B}$ is defined in eq.~\ref{B matrix }, the asymptotic rate constant becomes
\begin{equation}\label{SMD16}
    \begin{split}
     k^\infty _{mTGD}  &= 4a (2\pi \epsilon)^{(N-3)/2} \times \\
     & \times \frac 
     { D_{mTGD} \ \rho _{eq} (\mathbf{x}_s ) }
     { \sqrt {\det (\mathbf{B})  \det[ ({\mathbf{\Sigma}}^{-1})^{++}]   }   }.
     \end{split}
\end{equation}
Then by specifying the equilibrium distribution $\rho_{eq}(\mathbf{x}_s )$ at the saddle point according to the definition eq.~\ref{mTGD potential} of mTGD potential and by taking into account eq.~\ref{SMC8}, 
\begin{equation}\label{SMD17}
    \rho _{eq} (\mathbf{x}_s)=
    \frac{\exp(-\Delta U)}
   {2(2\pi)^{N/2} \sqrt { \det(\mathbf{\Sigma}) } }, 
\end{equation}
the final form of the asymptotic rate constant of mTGD model is recovered as reported by  eq.~\ref{mTGD potential} of the main text, where, in order to restore the original model, a unitary value has been attributed to parameter $\epsilon$ previously introduced only for recognizing the asymptotic limit

\newpage

\end{document}